\documentclass[]{article}
\usepackage[colorlinks]{hyperref}
\usepackage{framed}
\usepackage{graphicx}
\usepackage{appendix}
\usepackage{arydshln}
\usepackage{amssymb,amsfonts}
\usepackage{amsmath}
\usepackage{amsthm}
\usepackage{bbold}
\usepackage{caption}
\usepackage{fancyhdr}
\usepackage{geometry}
\usepackage{cite}
\usepackage{url}
\usepackage{babel}
\usepackage[utf8]{inputenc}\usepackage{babel}

\newcommand{\tr}{\mbox{$\mathrm{Tr}$}}
\newcommand{\lb}{\lbrace}
\newcommand{\rb}{\rbrace}

\title{Quantum generalized Calogero-Moser systems from free Hamiltonian reduction}
\author{Katarzyna Kowalczyk-Murynka\footnote{email: kkowalczyk@cft.edu.pl} \hskip 5pt and Marek Ku\'s\footnote{email: marek.kus@cft.edu.pl} \\
\textit{Center for Theoretical Physics, Polish Academy of Sciences} \\
\textit{Al. Lotnik\'ow 32/46, 02-668 Warszawa, Poland}}

\begin{document}
\renewcommand{\figurename}{Fig.}
\renewcommand{\tablename}{Tab.}

\maketitle

\begin{abstract}
The one-dimensional system of particles with a $1/x^2$ repulsive potential is known as the Calogero-Moser system. Its classical version can be generalised by substituting the coupling constants with additional degrees of freedom, which span the $\mathfrak{so}(N)$ or $\mathfrak{su}(N)$ algebra with respect to Poisson brackets. We present the quantum version of this generalized model. As the classical generalization is obtained by a symplectic reduction of a free system, we present a method of obtaining a quantum system along similar lines. The reduction of a free quantum system results in a Hamiltonian, which preserves the differences in dynamics of the classical system depending on the underlying, orthogonal or unitary, symmetry group. The orthogonal system is known to be less repulsive than the unitary one, and the reduced free quantum Hamiltonian manifests this trait through an additional attractive term $\sum_{i<j}\frac{-\hbar^2}{(x_i-x_j)^2}$, which is absent when one performs the straightforward canonical  Dirac quantization of the considered system. We present a detailed and rigorous derivation of the generalized quantum Calogero-Moser Hamiltonian, we find the spectra and wavefunctions for the number of particles $N=2,3$, and we diagonalize the Hamiltonian partially for a general value of $N$.  
\noindent
\end{abstract}

\section{Introduction}
 
 A one-dimensional system of $N$ mutually interacting particles described by the Hamiltonian
 \begin{equation}\label{CM}
 	\mathcal{H}_{CM}=\frac{1}{2}\sum_{i} p_i^2 + \frac{g^2}{2}\sum_{i\neq j}\frac{1}{(x_i-x_j)^2}, 
 \end{equation} 
 was proposed in its quantum version, (i.e.\ with $p_i$ and $x_i$ as canonical momentum and position operators) in a series of papers of F. Calogero  \cite{calogero69,calogero69a,calogero71}, where it was also shown 
 that the problem is soluble (integrable) by finding explicit expressions for the spectrum. He also conjectured \cite{calogero71} that the classical counterpart, where $p_i$ and $x_i$ are canonical phase-space variables, is completely integrable. Soon, it was rigorously shown by Moser \cite{moser75} and the system described by (\ref{CM}) is called the Caloger-Moser system. 
 
 The complete integrability of the considered system can be proved by exhibiting $N$ independent, involutive (i.e.\ with mutually vanishing Poisson brackets) integrals of motion, what was done in \cite{moser75}. The classical Arnold-Liouville theorem 
 leads then to the desired conclusion. 
 
 As shown by Kazhdan, Kostant, and Sternberg \cite{kazhdan78}, the complete integrability of the Calogero-Moser system can be established by employing the technique of the symplectic reduction described in a seminal paper of Marsden and Weinstein \cite{marsden74}. The general idea is to start from a Hamiltonian system in some larger phase space and then using symmetries, e.g, described by some group of canonical transformations, reduce the number of variables. The Marsden-Weinstein scheme guaranties that the reduced system will be again Hamiltonian in the reduced phase-space. Complete integrability will be inherited as well. In the considered 
 case we start form the free motion in an appropriate larger space with particular initial conditions. The reduction gives then the Calogero-Moser dynamics, which is thus obviously completely integrable since the free motion in the original phase space was such. 
 
The Calogero-Moser system can be generalized in a number of ways without losing the property of complete integrability (see \cite{kowalczykmurynka22} and the references cited therein). Firstly, one can add an external harmonic potential and a harmonic pair interaction potential \cite{calogero71}. Secondly, the pair interaction potential $ x^{-2} $ can be changed to $\sin^{-2}$ \cite{sutherland71,sutherland72},  
 $\sinh^{-2}$ \cite{calogero75} or the Weierstrass function $ \mathcal{P}(x) $ \cite{olshanetsky81}. More profound change consists in upgrading the coupling constant $g$ to a set of additional dynamical variables $ \{L_{ij}\} $ in an extended phase space (to be specified below)\footnote{In fact the "phase-space" in not, as it is usual, a symplectic manifold, but only a Poisson one, but its sufficient for all purposes of the paper} 
 with Poisson brackets 
 \begin{equation}\label{pbo}
 	 \left\{ {{L_{ij}},{L_{kl}}} \right\} = \frac{1}{2}\left( {{\delta _{kj}}{L_{il}} - {\delta _{il}}{L_{kj}} - {\delta _{jl}}{L_{ik}} + {\delta _{ik}}{L_{lj}}} \right),
 \end{equation}
 or
  \begin{equation}\label{pbu}
  \left\{ L_{ij},L_{kl}\right\} =\delta_{kj}L_{il}-\delta_{il}L_{kj}. 
  \end{equation}

 Poisson commuting with the canonical variables $ p_i, q_i $, $ 	\left\{ p_m,L_{kn}\right\} =\left\{ q_m,L_{kn}\right\}=0 $. The resulting generalized Calogero-Moser system 
 \begin{equation}\label{CML}
 	\mathcal{H}_{CM,L}=\frac{1}{2}\sum_{i} p_i^2 - \frac{1}{2}\sum_{i\neq j}\frac{L_{ij}L_{ji}}{(x_i-x_j)^2}, 
 \end{equation} 
 is completely integrable as can be shown by exhibiting appropriate number of integrals of motion \cite{wojciechowski85} or an appropriate reduction procedure \cite{gibbons84}. In fact, one can again apply the Marsden-Weinstein reasoning to the free motion in an extended phase-space. In the next section we will shortly show how to do it by explicit calculations without invoking general theorems.    
 
 The additional variables $ L_{ij} $ can be translated into some internal degrees of freedom of individual particles \cite{kowalczykmurynka22}. Hence, (\ref{CML}) becomes an interesting example of an integrable system of interacting particles with some internal structure. Moreover it found applications in investigations of spectral properties of quantum chaotic systems \textit{via} the so-called "level-dynamics" where $x_i$ are interpreted as eigenvalues of a quantum Hamiltonian and some perturbation parameter controlling degree of chaocity playing the role of time \cite{pechukas83}, \cite{yukawa85}, \cite{haake19}.
 
 As mentioned above, originally the Calogero-Moser Hamiltonian (\ref{CM}) ( to which we shall refer as to the ordinary CM system) was considered from the quantum point of view. It is appealing, thus, to look at the classical Hamiltonian of the generalized Calogero-Moser system from this point of view, i.e. to quantize it. A straightforward way to do it is to adopt the Dirac quantitation procedure by upgrading the phase-space variables to operators and Poisson brackets (\ref{pbo}) or (\ref{pbu}). The other method is suggested by the fact that the classical Hamiltonian (\ref{CML}) was obtained by reduction from the extended phase space. It should be thus possible to use some analogue of the classical reduction i.e., quantize (in this case the straightforwardly quantizable classical free Hamiltonian) and "project it" to the "reduced" Hilbert space using symmetries of the original system and separating the redundant variables. 
 
 The general scheme of such reduction was described in \cite{feher08}. In section \ref{SEC:QUANTRED}, again without invoking general theorems, we show by explicit calculations how the reduction works for the generalized Caloger-Moser system. We will show that, to some extend astonishingly, the two quantization procedures described above, i.e. the canonical Dirac quantization and the reduction one are not exactly equivalent and the discrepancy depends on the symmetry of the system, i.e.\ on the structure of the set of variables $ \{L_{ij}\} $ and the resulting form of the Poisson brackets (or, respectively, commutators) (\ref{pbo})  or (\ref{pbu}). 
    
\section{Classical reduction procedure}\label{sec:matrix_dyn}
As outlined in the introduction, the Calogero-Moser system  as well as the generalized ones can be obtained as a reduction of a linear Hamiltonian system. Here we only shortly outline the whole procedure, for details we refer to the reference \cite{kowalczykmurynka22} and the literature cited therein. First, we recall that a standard setting for Hamiltonian dynamics is that of a phase space, i.e., a differentiable manifold equipped with a symplectic form (a closed and non-degenerate two-form) and a Hamiltonian function determining dynamics \cite{abraham78}. In the most common case the phase space is the cotangent bundle $T^*\mathcal{M}$ of a configuration space $\mathcal{M}$. The construction simplifies if $\mathcal{M}$ is a linear space. Then, $T^*\mathcal{M}=\mathcal{M}\times\mathcal{M}$. 

We start by taking as a configuration space a linear space of  $N\times N$ complex Hermitian matrices: $\mathcal{M}=\{X\in M_N(\mathbb{C}): X^\dagger=X \}$ and the corresponding phase space $\mathcal{M}\times\mathcal{M}=\{(X,Y): X^\dagger=X,Y^\dagger=Y\}$, with the canonical symplectic form ,  
\begin{equation}\label{symplecticform}
	\omega=\sum_{i,j}dX_{ij}\wedge dY_{ji}=\tr(dX\wedge dY),
\end{equation}

The corresponding Poisson brackets\cite{abraham78} read,
\begin{equation}\label{PBmatr}
	\left\{f,g\right\}=\tr\left(\frac{\partial f}{\partial X}\frac{\partial g}{\partial Y}-
	\frac{\partial g}{\partial X}\frac{\partial f}{\partial Y}\right).	
\end{equation}

The evolution of an arbitrary phase-space function $f$ generated by a Hamiltonian function $\mathcal{H}$ is given by 
$\label{hameqs}
	\dot{f}=\left\{f,\mathcal{H}\right\}.
$

The Hamilton function $\label{HCM_mat}
	\mathcal{H}=\frac{1}{2}\tr YY^\dagger=\frac{1}{2}\tr Y^2$ generates the free motion $ \dot{X}=Y, \quad  \dot{Y}=0,$
i.e., 
\begin{equation} \label{lin_mat_dyn}
	X(t)=X_0+t\cdot Y_0, \quad Y(t)=Y_0.
\end{equation}
Now we choose a new time-dependent frame in which $X(t)$ remains diagonal,
\begin{equation} \label{diagonal}
	U(t)\left(\begin{array}{c} X(t) \\ Y(t) \end{array}\right)U(t)^\dagger =\left(\begin{array}{c} D(t) \\ V(t) \end{array}\right),
\end{equation}
with $D(t)$ diagonal. Denoting $A:=\dot{U}U^\dagger$ and $L:=[D,V]$ we obtain the following equations of motion  
\begin{equation}\label{DVL}
	\dot{D}=V+\left[A,D\right], \quad \dot{V}=\left[A,V\right], \quad \dot{L}=\left[A,L\right].  	
\end{equation}
Further, denoting $x_i=X_{ii}$, and $p_i=V_{ii}$ we arrive at the closed system of equations of motion for $x_i, p_i, L_{ij}$ (all remaining variables separate),
\begin{eqnarray}
	\dot{x}_{i}&=& p_{i} \label{xdot}\\
	\dot{p}_{i}&=& \sum_{k\neq i} \frac{-2L_{ik}L_{ki}}{(x_{i}-x_{k})^3} \label{pdot}\\ 
	\dot{L}_{ij}&=& \sum_{k \neq i,j} L_{ik}L_{kj}\left(\frac{1}{(x_{i}-x_{k})^2}-\frac{1}{(x_{j}-x_{k})^2}\right). \label{Ldot}
\end{eqnarray}

Equations (\ref{xdot})-(\ref{Ldot}) are Hamilton equations of motion derived from the old Hamilton function,  $\mathcal{H}=\frac{1}{2}\tr Y^2=\frac{1}{2} \tr V^2$ expressed in the new parametrisation, which happens to be exactly (\ref{CML}).
The Poisson brackets derived from (\ref{PBmatr}) take the form (\ref{pbu}). Since $L$ is anti-Hermitian we have $-L_{ij}L_{ji}=|L_{ij}|^2$, hence the interactions between particles are repulsive.

The same procedure applied to $H(X,Y)=\frac{1}{2}\tr(X^2+Y^2)$ leads to a similar system of repelling particles, but in an external harmonic potential \cite{olshanetsky81}, 
\begin{equation}
\mathcal{H}_{C,L}=\frac{1}{2}\sum_{i} \left(p_i^2+x_i^2\right) - \frac{1}{2}\sum_{i\neq j}\frac{L_{ij}L_{ji}}{(x_i-x_j)^2} \label{CL}
\end{equation}

If the initial matrices $(X_0,Y_0)$ are real and symmetric they will stay such during the whole evolution (\ref{lin_mat_dyn}).  This case, however, can be treated independently by choosing as the phase space $T^*\mathcal{M}=\mathcal{M}\times\mathcal{M}=\{(X,Y): X^T=X,Y^T=Y\}$, i.e., the cotangent bundle to the space of real symmetric matrices. The reduction procedure in this case is the same as presented above. The resulting equations are exactly (\ref{xdot})-(\ref{Ldot}), and the Hamilton function is the same as (\ref{CML}) (or (\ref{CL}) in the harmonic case). The only differences are that $X(t),Y(t)$ are real, symmetric matrices, $L(t)$ is real, antisymmetric and, instead of a unitary matrix $U(t)$, we use an orthogonal diagonalizing matrix $O(t)$, and the Poisson brackets take now the form (\ref{pbo}).

\section{Quantization of the generalized CM system}\label{SEC:QUANTRED}
\subsection{Canonical quantization}
The simplest approach towards a quantum Calogero-Moser system is to directly quantize (\ref{CML}) using the Dirac prescription:
\[L_{ij}\longrightarrow \hat{L}_{ij},\hspace{5mm}\left\{ {{L_{ij}},{L_{kl}}} \right\}\longrightarrow \frac{1}{i\hbar}\left[ {{\hat{L}_{ij}},{\hat{L}_{kl}}} \right],\]
where the Poisson brackets are given by (\ref{pbo}) or (\ref{pbu}) for $\alpha=1,2$ respectively. Restoring the $\frac{1}{m}$
factors for completeness, in the orthogonal setting,$\alpha=1$, we will obtain:
\begin{equation}
	\hat{H}^{(1)}_D= -\frac{\hbar^2}{2m} \sum_{i=1}^N\frac{\partial^2}{\partial x_i^2}-\frac{1}{2m}\sum_{i\neq i}\frac{\hat{L}^{(1)}_{ij}\hat{L}^{(1)}_{ji}}{(x_i-x_j)^2},
\end{equation} 
where $\hat{L}^{(1)}_{ij}/(i\hbar),i<j$ satisfy (\ref{pbo}) and $\hat{L}^{(1)}_{ji}=-\hat{L}^{(1)}_{ij}$.  In the unitary case we obtain:
\begin{equation}
	\hat{H}^{(2)}_D= -\frac{\hbar^2}{2m} \sum_{i=1}^N\frac{\partial^2}{\partial x_i^2}-\frac{1}{2m}\sum_{i\neq j}\frac{\hat{L}^{(2)}_{ij}\hat{L}^{(2)}_{ji}}{(x_i-x_j)^2},
\end{equation}
where $\hat{L}^{(2)}_{ij}/(i\hbar),i\neq j$ satisfy (\ref{pbu}). 

The $\hat{L}_{ij}$ operators can, of course, be realized by finite-dimensional matrices spanning the representations of the respective Lie algebras. This kind of models, with the corresponding Hilbert space $L^2(\mathbb{R}^N)\otimes \mathbb{C}^n$ (where $n$ is the dimension of the representation) have a somewhat unexpected property. In some cases, such as the $N$-dimensional, defining representations of $\mathfrak{so}(N)$ and $\mathfrak{su}(N)$, where:\[\hat{L}^{(1)}_{ij}=\frac{i\hbar}{2}\left(|i\rangle\langle j|-|j\rangle\langle i|\right), \hspace{5mm} \hat{L}^{(2)}_{ij}=i\hbar|j\rangle\langle i|,\]  the two Hamiltonians coincide up to a global factor in the interacting terms. On the other hand, as we show in our upcoming paper, the classical dynamics of the $L_{ij}$ degrees of freedom depend very strongly on the type of inbuilt symmetry. This motivates us to search for a model, which manifests the distinction between $\mathfrak{so}(N)$ and $\mathfrak{su}(N)$ models on the quantum level.   

\subsection{Reduction of a free quantum system}\label{freeQR}
An alternative approach towards quantization of (\ref{CML}), which we show to restore the difference between the $SO(N)$ and $SU(N)$ models, is based directly on the classical reduction procedure outlined in section \ref{sec:matrix_dyn}.
The classical generalisation of the $N$-particle Calogero-Moser model made use of the configuration space of Hermitian matrices:  $\mathcal{M}=\left\lbrace X\in M_{N\times N}(\mathbb{C}): X^{\dagger}=X \right\rbrace$. The phase space which is generally a cotangent bundle $T^*\mathcal{M}$ here is simply $M=\mathcal{M}\times\mathcal{M}$ consisting of pairs $(X,Y)$ of Hermitian matrices. Each such pair of matrices can be identified with a point in $\mathbb{R}^{2N^2}$, where  $1\leq i<j\leq N$ in the following way:
\begin{equation}
X = \left(\begin{array}{ccccc} x_{11} &...& ...& ...& ... \\
...    & x_{ii}& ... & \frac{x_{ij}^R+ix_{ij}^I}{\sqrt{2}}& ... \\
...&...&...&...&...\\
...&\frac{x_{ij}^R-ix_{ij}^I}{\sqrt{2}}& ...& x_{jj}&...\\
...&...&...&...& x_{NN}
\end{array}\right)\label{XinRN2},
\end{equation}
\begin{equation} 
Y = \left(\begin{array}{ccccc} y_{11} &...& ...& ...& ... \\
...    &y_{ii}& ... & \frac{y_{ij}^R+iy_{ij}^I}{\sqrt{2}}& ... \\
...&...&...&...&...\\
...&\frac{y_{ij}^R-iy_{ij}^I}{\sqrt{2}}& ...& y_{jj}&...\\
...&...&...&...& y_{NN} 
\end{array}\right) \label{YinRN2}
\end{equation}
The symplectic form in a $(x_{ii},x_{ij}^R,x_{ij}^I,y_{ii},y_{ij}^R,y_{ij}^I)\in\mathbb{R}^{2N^2}$ phase space:
\begin{equation}
\omega_R = \sum_{i=1}^N dx_{ii}\wedge dy_{ii}+\sum_{i<j}dx_{ij}^R\wedge dy_{ij}^R+dx_{ij}^I\wedge dy_{ij}^I
\end{equation}
coindices with $\omega = \tr(dX\wedge dY)$ for $(X,Y)$ defined with (\ref{XinRN2}) and (\ref{YinRN2}) (all the imaginary terms cancel). The Poisson brackets for $i<j,k<l$:
\begin{eqnarray}
\left\lbrace x_{ii},y_{jj}\right\rbrace =\delta_{ij},&  &\left\lbrace x_{ij}^{R,I},y_{kl}^{R,I}\right\rbrace = \delta_{ik}\delta_{jl} \\
\left\lbrace x_{ij}^R,x_{kl}^I\right\rbrace = 0,&  &\left\lbrace y_{ij}^R,y_{kl}^I\right\rbrace = 0 \\
\left\lbrace x_{ij}^R,y_{kl}^I\right\rbrace = 0,&  &\left\lbrace x_{ij}^I,y_{kl}^R\right\rbrace = 0
\end{eqnarray}
are equivalent to $\left\lbrace X_{ij},Y_{lk}\right\rbrace =\delta_{ik}\delta_{jl}$ in the matrix formulation. The variables $(\bar{x},\bar{y})\in\mathbb{R}^{2N^2}$ form a familiar ground for canonical quantisation. To each phase space variable we assign an operator acting on the Hilbert space $L^2(\mathbb{R}^{N^2})$:
\begin{eqnarray}
x_{ii}\longrightarrow \hat{x}_{ii},&  & y_{ii}\longrightarrow -i\hbar\partial_{x_{ii}} \\
x_{ij}^R\longrightarrow \hat{x}_{ij}^R,&  & y_{ij}^R\longrightarrow -i\hbar\partial_{x_{ij}^R} \\
x_{ij}^I\longrightarrow \hat{x}_{ij}^I,&   & y_{ij}^I\longrightarrow -i\hbar\partial_{x_{ij}^I} \\
\left\lbrace ... , ... \right\rbrace &\longrightarrow & \frac{1}{i\hbar}[...,...]
\end{eqnarray}

The operators $\hat{x}_{ij}^{R,I}$ and their canonical conjugates can be organised into matrices $\hat{X}$ and $\hat{Y}$ just as in (\ref{XinRN2}),(\ref{YinRN2}). Using such operator-valued matrices we can efficiently define the free Hamiltonian together with its eigenfunctions:
\begin{eqnarray}\label{freeN2}
\hat{H}_F &=& \frac{1}{2m}\tr\left(\hat{Y}^2\right),\hspace{5mm}\psi_K(X)=e^{i\tr(KX)},\\
\hat{H}_F\psi_K(X)&=& E_K\psi_{K}(X),\hspace{5mm} E_K=\frac{\hbar^2}{2m}\tr\left(K^2\right), 
\end{eqnarray}
where $K$ is a Hermitian matrix. Following the idea of classical reduction presented in section \ref{sec:matrix_dyn} We introduce new variables $(D,\bar{a})\in \mathbb{R}^{N+d}$ where $D$ is a diagonal matrix, $U=\exp(\bar{a}\cdot\bar{\tau})$ is unitary and $\bar{\tau}=(\tau_1,\tau_2,...,\tau_d)$ are the anti-Hermitian basis vectors of the Lie algebra $\mathfrak{su}(N)$:
\begin{equation}
X = U^{\dagger}DU=\exp(-\bar{a}\cdot\bar{\tau})D\exp(\bar{a}\cdot\bar{\tau}).
\end{equation}
In the orthogonal setting the $(x^I,y^I)$ degrees of freedom are of-course redundant, the off-diagonal elements of the real, symmetric matrices $(X, Y)$ are simply $(x,y)^R_{ij}$, $K$ is real and symmetric as well, and $\bar{\tau}=(\tau_1,\tau_2,...,\tau_d)$ are the anti-symmetric basis vectors of the Lie algebra $\mathfrak{so}(N)$. In this notation most of the expressions will be valid in both orthogonal and unitary case. 

The Hamiltonian and the wavefunction (\ref{freeN2}) can be re-expressed in these variables:
\begin{eqnarray}
\psi_K(D,\bar{a})&=&\exp(i\tr(\exp(\bar{a}\cdot\bar{\tau})K\exp(-\bar{a}\cdot\bar{\tau})D))=e^{i\tr(K(a)D)}, \label{waveN2} \\
\hat{H}_{D,\bar{a}}&=&-\frac{\hbar^2}{2m}\sum_{i,j,k=1}^N\sum_{l=1}^d\left(\frac{\partial D_k}{\partial X_{ij}}\frac{\partial}{\partial D_k}+\frac{\partial a_l}{\partial X_{ij}}\frac{\partial}{\partial a_l} \right)^2, \label{HfreeDa}
\end{eqnarray}
We will prove that $H_{D,\bar{a}}$ differs by a similarity transformation from the generalized Calogero-Moser Hamiltonian with $(D_i-D_j)^{-2}\hat{\lambda}^2_{ij}$ interacting terms, where $\hat{\lambda}_{ij}=\sum_{k=1}^d l_{ij}^k\partial_{a_k}$, and the commutators are the exact counterparts of the classical Poisson brackets (\ref{pbo}) or (\ref{pbu}). The similarity transformation depends on the symmetry, and turns out to produce additional interacting terms, which distinguish the orthogonal and unitary Hamiltonian. The last step of the reduction procedure will consist, whenever it is possible, of projecting $\psi_K(D,\bar{a})$ on the  $\hat{\lambda}_{ij}$ eigenspaces in a way that will result in eigenfunctions of the CM Hamiltonian. 

Before carrying out this programme we need to make an important remarks. As already mentioned, if we start with real, symmetric matrices $X$ and $K$, we have $U\in SO(N)$ and $(\tau_1,\tau_2,...,\tau_d)$ spanning the $\mathfrak{so}(N)$ Lie algebra. In this case the configuration space $\mathcal{M}\equiv \mathbb{R}^{N+\binom{N}{2}}$, $N+d=N+\binom{N}{2}$, and there are as many generators and corresponding $(a_1,...,a_d)$ variables as needed. On the other hand in the unitary case $\mathcal{M}\equiv \mathbb{R}^{N^2}$ while the dimension of $su(N)$ is equal to $N^2-1$, so it seems we have $N+d=N^2+N-1$ and $X\longrightarrow (D,a_1,...,a_d)$ would not be a valid coordinate transformation. Fortunately we can restrict ourselves to unitary matrices with real diagonal entries due to gauge equivalence: if some matrix $U$ diagonalises a given $X$, so does $U'=diag(e^{i\phi_1},...,e^{i\phi_N})U$, and if $U'_{ii}\in \mathbb{R}$, $U'$ can be generated by $N^2-N$ off-diagonal generators only. Therefore we can set $d=N^2-N$ and stay in the off-diagonal $\mathfrak{su}(N)$ subspace. 

\subsection{The free Hamiltonian in $(D,a)$ variables}\label{DaH}
The $X_{ij}$ variables can be eliminated from (\ref{HfreeDa}) \textit{via} the following expressions:
\begin{equation}
X_{ij}(D,\bar{a})= (U^{\dagger}(\bar{a})DU(\bar{a}))_{ij},\hspace{5mm}U = U(\bar{a}) = e^{\bar{a}\bar{\tau}}.
\end{equation}
The derivatives $\partial_{X_{ij}}D_k$ and $\partial_{X_{ij}}a_l$ can be found as the elements of the inverted Jacobian matrix:
\begin{equation}\label{JacXDa}
\frac{\partial(D,\bar{a})}{\partial X}=\left(\frac{\partial X}{\partial(D,\bar{a})}\right)^{-1}=\left(\frac{\partial (U^{\dagger}(\bar{a})DU(\bar{a})))}{\partial(D,\bar{a})}\right)^{-1},
\end{equation}
\begin{equation}\label{Jac}
\frac{\partial X}{\partial(D,\bar{a})}=\left(\begin{array}{c : c}
\frac{\partial X_{ii}}{\partial D_k}=|U_{ki}|^2 \hspace{2mm}&\hspace{2mm} \frac{\partial X_{ii}}{\partial a_l}=(U^{\dagger}\Omega_lU)_{ii}\\
\\
\hdashline
\\
\frac{\partial X_{ij}}{\partial D_k}=U_{ki}^*U_{kj} \hspace{2mm} & \hspace{2mm} \frac{\partial X_{ij}}{\partial a_l}=(U^{\dagger}\Omega_lU)_{ij} \\
\\
\end{array}\right),
\end{equation}
where 
\begin{eqnarray}\label{omegas}
\Omega_l&=&\left[D,(\partial_{a_l}U)U^{\dagger}\right]=\Omega_l^{\dagger},\\
(\partial_{a_l}U)U^{\dagger}&=& u(A)_{lk}\tau_k\in\mathfrak{g}, \\
u(A)&=&\sum_{n=0}^\infty \frac{A^n}{(n+1)!}, \hspace{5mm}A_{ij}=a_k\cdot f_{ki}^j
\end{eqnarray} 
Now the challenging part is to invert this matrix. We can also use it to calculate the metric tensor\footnote{i.e. the Euclidean metric tensor on $\mathbb{R}^{N+d}$, which is the identity matrix in the $X$ coordinates, expressed \emph{via} $(D,a)$ coordinates}, as shown in the appendix \ref{derivDaH}:
\begin{equation}\label{gDa}
\mathbf{g}=\left(\begin{array}{c : c}
\mathbb{1}_{N\times N}& 0_{N\times d} \\

\hdashline
\\
0_{d\times N}  & \tr(\Omega_l\Omega_k)
\\
\end{array}\right),
\end{equation}
and then, due to the block structure of $\mathbf{g}=\mathbf{1}_N\oplus g$, we clearly see that the Hamiltonian splits into two parts:
\begin{eqnarray}\label{NDaHamilt}
\hat{H}&=&-\frac{\hbar^2}{2m}\frac{1}{\sqrt{|\mathrm{det} g|}}\left(\sum_{k=1}^N\partial_{D_k}\sqrt{|\mathrm{det} g|}\partial_{D_k}+\sum_{k,l=1}^d\partial_{a_l}\sqrt{|\mathrm{det} g|}(g^{-1})_{lm}\partial_{a_m}\right)\nonumber\\
&=&-\frac{\hbar^2}{2m}(\Delta_D+\Delta_a)
\end{eqnarray}
For both orthogonal and unitary matrices we prove in (\ref{derivDaH}) that the matrix elements of $g$ have the form:
\[ g_{(ij)(kl)}=\frac{1}{2}\sum_{(pq)\in I}u_{(ij)(pq)}(D_p-D_q)^2(u^T)_{(pq)(kl)}=\frac{1}{2}(u \bold{D}^2 u^T)_{(ij)(kl)},
\]
and therefore
\begin{equation}
g = \frac{1}{2}u \bold{D}^2 u^T,\hspace{5mm}\bold{D}=diag(D_i-D_j,(ij)\in I)\in M_{d\times d}\label{g}
\end{equation}
where $I$ is the set of indices appropriate for each setting. In the orthogonal case $I=\lb (ij): 1\leq i < j \leq N \rb$ and in the unitary case $I=\lb (ij): 1\leq i \neq j \leq N \rb$. From this it automatically follows that:
\begin{eqnarray}
\mathrm{det} g&=&2^{-d}\mathcal{D}^2|\mathrm{det} u|^2,\hspace{5mm}\mathcal{D}=\prod_{i<j}|D_i-D_j|^{\alpha},\label{detg}\\
g^{-1}&=&2(u^T)^{-1}\bold{D}^{-2}u^{-1}\label{g1}
\end{eqnarray}
where $\alpha=\frac{2d}{N(N-1)}$, $\alpha=1$ corresponds to the orthogonal and $\alpha=2$ to the unitary case. The factorisation of $g$ and $\mathrm{det} g$ into a purely $D$ and $\bar{a}$-dependent parts allows us to simplify $\Delta_D$ (appendix \ref{derSIMD}):
\begin{eqnarray}
\Delta_D&=&\frac{1}{\mathcal{D}}\sum_{i=1}^N\frac{\partial}{\partial D_i}\left(\mathcal{D}\frac{\partial}{\partial D_i}\right)\\
\sqrt{\mathcal{D}}\Delta_D\frac{1}{\sqrt{\mathcal{D}}}&=&\sum_{i=1}^N\frac{\partial^2}{\partial^2 D_i}+\frac{\alpha(2-\alpha)}{2}\sum_{i<j}^N\frac{1}{(D_i-D_j)^2}\label{SIMD},
\end{eqnarray}
where (\ref{SIMD}) is a similarity transformation which eliminates the $1^{st}$ order derivatives from $\Delta_D$ (and leaves the $\Delta_a$ part unchanged). The result is a free $N$ particle system for the unitary and an ordinary CM for the orthogonal case\footnote{We notice that the interaction term is attractive in this case, but this is not the full Hamiltonian, and further repulsive terms will appear.}.

Before we move on to $\Delta_a$, let us recall that the indices of the algebra degrees of freedom are ordered pairs $(ij),(pq),(rs)\in I$, where $I=\lb(pq): 1\leq p < q \leq N\rb$ in the orthogonal case and $I=\lb(pq): 1\leq p \neq q \leq N\rb$ in the unitary case. Now we can write $\Delta_a$ explicitly:
\begin{eqnarray}\label{Deltaa}
\Delta_a&=&\frac{2}{|\mathrm{det} u|}\sum_{(ij)\in I}\sum_{(pq)\in I}\sum_{(rs)\in I}\frac{\partial}{\partial a_{pq}}\left(|\mathrm{det} u|\frac{u^{-1}_{(ij)(pq)}u^{-1}_{(ij)(rs)}}{(D_i-D_j)^2}\frac{\partial}{\partial a_{rs}}\right) \\
&=& 2\sum_{(ij)\in I}\frac{\hat{\Lambda}_{ij}}{(D_i-D_j)^2}
\end{eqnarray}

The free Hamiltonian, after similarity transformation, in the $(D,a)$ variables:
\begin{equation}\label{freeCMH}
\hat{H}'_F=\sqrt{\mathcal{D}}\hat{H}_F\frac{1}{\sqrt{\mathcal{D}}}=-\frac{\hbar^2}{2m}\left(\sum_{i=1}^N \frac{\partial^2}{\partial D_i^2}+\sum_{(ij)\in I}\frac{2\hat{\Lambda}_{ij}+(1-\frac{\alpha}{2})}{(D_i-D_j)^2}\right),
\end{equation}
has a form of a generalised Calogero-Moser Hamiltonian, where:
\begin{eqnarray}
\hat{\Lambda}_{ij}&=&\hat{\lambda}_{ij}^2+F_{ij}\hat{\lambda}_{ij} \label{Lambda_ij}\\
\hat{\lambda}_{ij}&=&\sum_{(pq)\in I}u^{-1}_{(ij)(pq)}\frac{\partial}{\partial a_{pq}} \\
F_{ij} &=&\frac{1}{|\mathrm{det} u|}\sum_{(pq)\in I}\frac{\partial}{\partial a_{pq}}\left(|\mathrm{det} u| u^{-1}_{(ij)(pq)}\right)=\frac{1}{|\mathrm{det} u|}\sum_{(pq)\in I}\left(\frac{\partial C(u)}{\partial a_{pq}}\right)_{(pq)(ij)},\label{F_ij}
\end{eqnarray}
where $C(u)$ is the cofactor matrix of $u$, and the determinant $\mathrm{det} u$ is positive, so the absolute value is redundant. Using the definition of a divergence of a tensor field, we can rewrite (\ref{F_ij}) as:
\begin{equation}
F_{ij}=\frac{1}{\mathrm{det} u}\left(\nabla\cdot C(u^T)\right)_{ij}.
\end{equation}
In the appendix \ref{derL} We prove that $F_{ij}$ vanishes identically. This means that indeed 
\begin{equation}\label{LAMBDA2}
\hat{\Lambda}_{ij}=\hat{\lambda}^2_{ij}
\end{equation}
and the last step is to calculate the commutation relations between the $\hat{\lambda}_{ij}$ operators. As shown in \ref{derL}, they turn out to be the following:
\begin{equation}\label{LIJKL}
\left[\hat{\lambda}_{ij},\hat{\lambda}_{kl}\right]=-\sum_{(mn)\in I} f_{(ij)(kl)}^{(mn)}\hat{\lambda}_{mn},
\end{equation}
which means that they form a representation of the Lie algebra $\mathfrak{so}(N)$ or $\mathfrak{su}(N)$ depending on the setting. 
This result is probably as close to quantizing the generalized Calogero-Moser model as one can get. The next step is to find  the eigenfunctions through appropriate integration of the plane waves (\ref{waveN2}).
 
 Finally, since $\tr(X^2)=\tr(D^2)$, adding an external harmonic potential to $\hat{H}_F$ will result in the same harmonic term added to the  Hamiltonian:
\begin{equation}\label{harmCMH}
\hat{H}'_H=\sqrt{\mathcal{D}}\hat{H}_H\frac{1}{\sqrt{\mathcal{D}}}=\sum_{i=1}^N\left(-\frac{\hbar^2}{2m} \frac{\partial^2}{\partial D_i^2}+\frac{m\omega^2}{2}D_i^2\right)-\frac{\hbar^2}{2m}\left(\sum_{(ij)\in I}\frac{2\hat{\Lambda}_{ij}+(1-\frac{\alpha}{2})}{(D_i-D_j)^2}\right).
\end{equation}

\subsection{The Hamiltonian and the reduced wave functions for N=2}\label{HRedN2}
We expect the case of $N=2$ particles to be equivalent to the ordinary CM system. With only one pair of interacting particles, there is by default only one operator $\hat{L}_{12}$ whose eigenvalue should yield a constant coupling in the interaction term $\frac{g^2}{(D_1-D_2)^2}$. This is why we will treat the $N=2$ (orthogonal and unitary) case separately. We shall apply the results from \ref{DaH} and calculate the reduced wave functions in the orthogonal and unitary case.
\subsubsection{SO(2)}
This example is not entirely an original contribution, similar considerations can be found in \cite{CARIENA2007}. The $SO(2)$ group has only one generator and it is known that its every element can be expressed as:
\begin{equation}
U(\phi)=\exp\left(\phi\left(\begin{array}{c c} 0 & -1 \\ 1 & 0

\end{array}\right)\right)=\left(\begin{array}{c c}\cos\phi & -\sin\phi \\ 
\sin\phi & \cos\phi
\end{array}\right).
\end{equation}
Since there is only one generator, there are no nonvanishing structure constants, and therefore in the adjoint representation $A=0$ and $u(A)=u^{-1}(A)=1$. The single $\Lambda$ operator at hand is $\hat{\Lambda}_{12}=\frac{\partial^2}{\partial\phi^2}$, and the Hamiltonian (\ref{freeCMH}) has a simple form:
\begin{equation}
\hat{H}'=-\frac{\hbar^2}{2m}\left[\frac{\partial^2}{\partial D_1^2}+\frac{\partial^2}{\partial D_2^2}+\frac{1}{2(D_1-D_2)^2}\left(\frac{\partial^2}{\partial\phi^2}+1\right)\right]
\end{equation}
which easily translates to centre of mass and relative distance variables $R=\frac{D_1+D_2}{2}$, $r=D_1-D_2$ and a doubled angle $\Phi=2\phi$:
\begin{equation}
\hat{H}'(R,r,\Phi)=-\frac{\hbar^2}{4m}\frac{\partial^2}{\partial R^2}-\frac{\hbar^2}{m}\left[\frac{\partial^2}{\partial r^2}+\frac{1}{r^2}\left(\frac{\partial^2}{\partial \Phi^2}+\frac{1}{4}\right)\right].
\end{equation}
Now we need to express the eigenfinction of the unreduced, free system in these variables:
\begin{eqnarray}
\psi_K(X)&=&\exp(i\tr(KX))=\exp\left[i\tr\left(KU^T(\phi)DU(\phi)\right)\right]=\psi_{K}(D,\phi)\nonumber\\
&=&\exp\left[i\tr\left(\left(\begin{array}{c c}K_1 & \frac{K_{12}}{\sqrt{2}} \\ 
\frac{K_{12}}{\sqrt{2}}  & K_2
\end{array}\right)U(\phi)^T\left(\begin{array}{c c}R+\frac{r}{2} & 0 \\ 
0 & R-\frac{r}{2}
\end{array}\right)U(\phi)\right)\right]\nonumber\\
\psi_K(R,r,\Phi)&=&\exp[[i\tr(KR)]\exp[i\kappa\cos(\Phi+\phi_k)r], \\
\kappa&=&\frac{1}{2}\sqrt{(K_1-K_2)^2+2K_{12}^2}=\sqrt{\tr(k^2)},\\
k&=& K-\frac{1}{2}\tr(K)\\
\cos\phi_k&=&\frac{K_1-K_2}{2\kappa}.
\end{eqnarray}
Now we make use of the fact, that $\hat{H}\psi_K=E_K\psi_K$ holds no matter in which variables we express the equation in, and also if we reverse the similarity transformation $\hat{H}'=\sqrt{r}\hat{H}\frac{1}{\sqrt{r}}$, we get:
\begin{equation}\label{freeCMSO2}
\hat{H}'(R,r,\Phi)(\sqrt{r}\psi_K(R,r,\Phi))=E_K\sqrt{r}\psi_K(R,r,\Phi).
\end{equation}
The crucial step of the  procedure is the integration over the $\Phi$ variable which projects $\psi_K(R,r,\Phi)$ onto an eigenspace of $\hat{\Lambda}_{12}=\partial^2_{\Phi}$. We perform the integration of both sides of (\ref{freeCMSO2}):
\[
-\frac{\hbar^2}{m}\int e^{-i\nu\Phi}\left[\frac{1}{4}\frac{\partial^2}{\partial R^2}+\frac{\partial^2}{\partial r^2}+\frac{1}{r^2}\left(\frac{\partial^2}{\partial \Phi^2}+\frac{1}{4}\right)\right](\sqrt{r}\psi_K)d\Phi= \newline E_K\sqrt{r}\int e^{-i\nu\Phi}\psi_Kd\Phi \]

where the integration cuts out a Fourier component of $\psi_{K}(R,r,\Phi)=\sum_{\nu\in\mathbb{Z}} e^{i\nu\Phi}\psi_{n,K}(R,r)$. The integral $\int_0^{2\pi} e^{-i\nu\phi}\psi_K(R,r,\Phi)d\Phi=\psi_{\nu,K}(R,r)$ commutes with the $R$ and $r$-dependent terms of $\hat{H}$, thus:
\begin{equation}
-\frac{\hbar^2}{m}\left(\frac{1}{4}\frac{\partial^2}{\partial R^2}+\frac{\partial^2}{\partial r^2}+\frac{1}{4r^2}\right)(\sqrt{r}\psi_{\nu,K})-\frac{\hbar^2}{mr^2}\sqrt{r}\int_0^{2\pi}e^{-i\nu\Phi}\frac{\partial^2\psi_{K}}{\partial \Phi^2}d\Phi=  E_K\sqrt{r}\psi_{\nu,K}\nonumber,
\end{equation}
where the only nontrivial term can be calculated by parts as for $\nu\in\mathbb{Z}$ the boundary terms cancel: 
\begin{equation}\int_0^{2\pi}e^{-i\nu\Phi}\partial^2_ {\Phi}\left(\psi_{K}(R,r,\Phi)\right)d\Phi=-\nu^2\int_0^{2\pi} e^{-i\nu\Phi}\psi_K(R,r,\Phi)d\Phi=-\nu^2\psi_{\nu,K}(R,r).
\end{equation}
Finally, we discover that the function $\sqrt{r}\psi_{\nu,K}(R,r)$ is the eigenfunction od an ordinary 2-particle Calogero-Moser Hamiltonian with $g=\nu^2-\frac{1}{4}$:
\begin{equation}
-\frac{\hbar^2}{m}\left[\frac{1}{4}\frac{\partial^2}{\partial R^2}+\frac{\partial^2}{\partial r^2}+\frac{1}{r^2}\left(\frac{1}{4}-\nu^2\right)\right](\sqrt{r}\psi_{\nu,K}(R,r))=  E_K\sqrt{r}\psi_{\nu,K}(R,r).
\end{equation}
In the final step we need to compute the integral:
\begin{eqnarray}
\psi_{\nu,K}(R,r)&=& e^{i\tr(K)R}\sqrt{r}\int_{0}^{2\pi}e^{-i\nu\Phi}e^{i\kappa\cos(\Phi+\phi_K)r}d\Phi=\\
&=& e^{i\nu\phi_K}e^{i\tr(K)R}\int_{0}^{2\pi}e^{i(\kappa r\cos\Phi-\nu\phi)}d\Phi= \nonumber\\&=& e^{i\nu\phi_K}e^{i\tr(K)R}i^{\nu} \sqrt{r}J_{\nu}(\sqrt{\tr(k^2)} r).
\end{eqnarray}
As expected \cite{calogero69}, the Calogero-Moser wave function, apart from the trivial center of mass component and some phase factors, has the form $\sqrt{r}J_{\nu}(\kappa r)$, where $J_{\nu}$ is the Bessel function. Moreover, the trace of $K$ contributes to the energy of center of mass motion, $E_{CM}=\frac{\hbar^2 \tr(K)^2}{4m}$, while the traceless part $k=K-\frac{1}{2}\tr(K)$ contributes to the relative motion $E_{rel}=\frac{\hbar^2}{m}\tr(k^2)$, $E_{CM}+E_{rel}=E_K$.

\subsubsection{SU(2)}
In the unitary case two parameters are necessary to define a diagonalising matrix (the third parameter is redundant as stated in \ref{freeQR}):
\begin{equation}
U\left(\theta,\phi\right)=\left(\begin{array}{c c}\cos\theta & -\sin\theta e^{-i\phi} \\ 
\sin\theta e^{i\phi} & \cos\theta
\end{array}\right),
\end{equation}
where $\theta\in\left[0,\pi/2\right]$ is a sufficient set, due to gauge symmetry. In order to find the metric tensor we need to calculate:
\begin{eqnarray*}
	(\partial_{\theta}U)U^{\dagger}&=&\left(\begin{array}{c c}0 & -e^{-i\phi} \\ e^{i\phi} & 0 \end{array}\right),\hspace{5mm}(\partial_{\phi}U)U^{\dagger}=\left(\begin{array}{c c}-i\sin^2\theta & i\sin\theta\cos\theta e^{-i\phi} \\ i\sin\theta\cos\theta e^{i\phi} & i\sin^2\theta \end{array}\right) \\
	\Omega_{\theta}&=& r\left(\begin{array}{c c}0 & -e^{-i\phi} \\ -e^{i\phi} & 0 \end{array}\right) ,\hspace{5mm} \Omega_{\phi}=r\left(\begin{array}{c c}0 & -i\sin\theta\cos\theta e^{-i\phi} \\ i\sin\theta\cos\theta e^{i\phi} & 0 \end{array}\right)
\end{eqnarray*}
where $r=D_1-D_2$. The nontrivial block of the metric tensor, its determinant and in:verse have the following form:
\begin{equation}
g=\left(\begin{array}{c c} 2r^2 & 0 \\ 
0 & \frac{1}{2}r^2\sin^2(2\theta)
\end{array}\right),\hspace{5mm}g^{-1}=\left(\begin{array}{c c} \frac{1}{2r^2} & 0 \\ 
0 & \frac{2}{r^2\sin^2(2\theta)}\end{array}\right), \hspace{5mm}\sqrt{|\mathrm{det}(g)|}=r^2\sin(2\theta).
\end{equation}
The $(\theta,\phi)$ dependent part of the Laplacian, after all the simplifications,and substitution $2\theta\longrightarrow \theta\in\left[0,\pi\right]$:
\begin{eqnarray}
\Delta_{\theta,\phi}&=&\frac{1}{r^2\sin(2\theta)}\left[\frac{\partial}{\partial\theta}\left(\frac{\sin(2\theta)}{2}\frac{\partial}{\partial\theta}\right)+\frac{2}{\sin(2\theta)}\frac{\partial^2}{\partial\phi^2}\right]=\nonumber\\
&=&\frac{2}{r^2}\left[\frac{1}{\sin\theta}\frac{\partial}{\partial\theta}\left(\sin\theta\frac{\partial}{\partial\theta}\right)+\frac{1}{\sin^2\theta}\frac{\partial^2}{\partial\phi^2}\right]=\frac{2}{r^2}\Delta_{S^2}
\end{eqnarray}
turns out to be proportional to a familiar Laplace operator on the two-dimensional sphere. The eigenfunction $\psi_{K}(X)=\psi_K(R,r,\theta,\phi)$ in the new variables:
\begin{eqnarray}
\psi_K(R,r,\theta,\phi)&=& e^{\left(i\tr(KU^{\dagger}DU)\right)}
= e^{i(K_1+K_2)R}\exp\left[i\left(\kappa_1\cos\theta-\kappa_2\sin\theta\right)r\right]\nonumber\\
K &=& \left(\begin{array}{c c}K_1 & \frac{K_R+iK_I}{\sqrt{2}} \\
\frac{K_R-iK_I}{\sqrt{2}} & K_2
\end{array}
\right)\\
\kappa_1 &=& \frac{K_1-K_2}{2} \\
\kappa_2 &=& \frac{K_R\cos\phi-K_I\sin\phi}{\sqrt{2}} =|K_{12}|\cos(\phi+\phi_k)\\
\cos\phi_k&=&\frac{K_R}{\sqrt{K_R^2+K_I^2}}
\end{eqnarray}
decomposes into spherical harmonics:
\begin{eqnarray}
\psi_K(R,r,\theta,\phi)&=&\sum_{l,m}Y_{l,m}(\theta,\phi)\psi_{l,m,K}(R,r) \\
\psi_{l,m,K}(R,r)&=&\int_{S^2}Y^*_{l,m}(\theta,\phi)\psi_{K}(R,r,\theta,\phi)\sin\theta d\theta d\phi.
\end{eqnarray}
We use this decomposition in a similar way as in the orthogonal case, only this time $\sqrt{\mathcal{D}}=r$:
\begin{equation}
\hat{H}\psi_K= E_K\psi_K,\hspace{5mm}\hat{H}'(r\psi_K)= E_K(r\psi_K),\hspace{5mm}\int Y^*_{l,m}\hat{H}'(r\psi_K)= E_K\int Y^*_{l,m}(r\psi_K),
\end{equation}
and since we can integrate by parts: $\int Y^*_{l,m}(\Delta_{S^2}\psi_K)=\int(\Delta_{S^2}Y^*_{l,m})\psi_K=-l(l+1)\int Y^*_{l,m}\psi_K$, we obtain a Calgero-Moser Hamiltonian with a coupling constant $l(l+1)$:
\begin{equation}
\left[-\frac{\hbar^2}{4m}\frac{\partial^2}{\partial R^2}+\frac{\hbar^2}{m}\left(-\frac{\partial^2}{\partial r^2}+\frac{l(l+1)}{r^2}\right)\right]r\psi_{l,m,K}= E_K r\psi_{l,m,K}.
\end{equation}
As previously, the final step is to calculate the reduced function: 
\begin{eqnarray*}
	\psi_{l,m,K}(R,r)&=&\int\int Y^*_{l,m}(\theta,\phi)\psi_K(R,r,\theta,\phi)\sin\theta d\theta d\phi =\\
	&=& e^{i\tr(K)R}\int\int P_l^m(\cos\theta)e^{-im\phi}\exp\left[i\left(\kappa_1\cos\theta-\kappa_2\sin\theta\right)r\right]\sin\theta d\theta d\phi = \\
	&=& e^{i\tr(K)R}\int P_l^m(\cos\theta)e^{i\kappa_1\cos\theta r}\left(\int e^{-im\phi-i|K_{12}|\sin\theta\cos(\phi+\phi_k)r}d\phi\right)\sin\theta d\theta  = \\
	&=&e^{i\tr(K)R+im\phi_k}\int P_l^m(\cos\theta)e^{i\kappa_1\cos\theta}J_m(|K_{12}|r\sin\theta)\sin\theta d\theta 
\end{eqnarray*}
The relative factor contains the Bessel function which arises from the integration over $\phi$. The integral over $\theta$ looks complicated, but an analytical solution exists \cite{Neves_2006}, and has the form:
\begin{equation}
I_l^m(K,r)\propto P_l^m\left(\frac{2\kappa_1}{\kappa}\right)j_l(\kappa r),
\end{equation} 
where $\kappa=\sqrt{\tr{(k^2)}}$ and $k$ is the traceless part of $K$, just like in the orthogonal case, and $j_l(x)$ is the spherical Bessel function defined as:
\begin{equation}
j_l(x)=\sqrt{\frac{\pi}{2x}}J_{l+\frac{1}{2}}(x).
\end{equation} 
This means that the inverse square root from $j_l$ and the $r$ factor from the similarity transformation simplify, and as it should be expected from the two-particle Calogero-Moser system with $g=\frac{\hbar^2}{4m}l(l+1)$, the relative wavefunction is $\sqrt{r}J_{l+\frac{1}{2}}(r)$.

To summarize, for $N=2$ the reduction of a free system results in ordinary Calogero-Moser systems with quantized values of coupling constants. The orthogonal setting recovers the case of $g=\frac{\hbar^2}{4m}\left(l^2-\frac{1}{4}\right)$ and the relative wavefunctions in the form of $\psi_l(r)=\sqrt{r}J_l(\kappa r)$, and the unitary one recovers $g=\frac{\hbar^2}{4m}l\left(l+1\right)$ and $\psi_{l+\frac{1}{2}}(r)=\sqrt{r}J_{l+\frac{1}{2}}(\kappa r)$.

\subsection{The Hamiltonian and reduced wavefunctions for $N \geq 3$}
For the simplest case of $N=2$ it was possible to compute the metric tensor and the Laplace operator directly. For $N\geq 3$ we have the general Hamiltonian:
\begin{eqnarray}
\hat{H}'_F&=&-\frac{\hbar^2}{2m}\sum_{i=1}^N \frac{\partial^2}{\partial D_i^2}+\frac{1}{2m}\sum_{(ij)\in I}\frac{2\hat{L}^2_{ij}+\hbar^2(\frac{\alpha}{2}-1)}{(D_i-D_j)^2}\label{hamilt}\\
\hat{L}_{ij}&=&-i\hbar\hat{\lambda}_{ij},\hspace{5mm}\left[\hat{L}_{ij},\hat{L}_{kl}\right]=\sum_{(mn)}i\hbar f_{(ij)(kl)}^{(mn)}\hat{L}_{mn}\\
\hat{L}^{\dagger}_{ij}&=& (-i\hbar \hat{\lambda}_{ij})^{\dagger}=i(-\hbar\hat{\lambda}_{ij})=\hat{L}_{ij}, 
\end{eqnarray}
where, as obtained \emph{via} $N\times N$ matrix diagonalization, the $\hat{L}$ operators are automatically, by construction, represented with $N\times N$ matrices as well (that is in the defining representation of $\mathfrak{so}(N)$ or $\mathfrak{su}(N)$). The squared operators $\hat{L}_{ij}^2$ present in the Hamiltonian happen to commute thus diagonalize simultaneously in this special case. On the other hand, nothing prevents us from considering the above model with the $\hat{L}$ operators expressed in other representations of the corresponding Lie algebras.  Regardless of the chosen representation, having more than one interacting pair of particles and more than a single $\hat{L}$ operator makes it difficult to produce eigenfunctions of the above Hamiltonian from plane waves defined in the $X$ variables. The reason for this is that if they do not commute, the spatial and internal variables will not separate, and even if they do commute, as in the case of the defining representation, they do not form a complete set of commuting observables. Thus we need a different approach than a simple projection which worked for $N=2$. The wave-function, which had the form:
\begin{eqnarray}
\sqrt{\mathcal{D}}\psi_K(D,a)&=&\sqrt{\prod_{i<j}(D_i-D_j)^{\alpha}}\cdot e^{i\tr(K(a)D)},\\
\hat{H}_F'(\sqrt{\mathcal{D}}\psi_K(D,a))&=& E_K(\sqrt{\mathcal{D}}\psi_K(D,a))
\end{eqnarray}
prior to the reduction, now will be a spinor of the appropriate dimension. Still, the resulting probability distribution, \[|\sqrt{\mathcal{D}}\psi_K(D,a)|^2=\prod_{i<j}(D_i-D_j)^{\alpha},\] where $\alpha=1,2$ in the orthogonal and unitary setting
respectively, resembles the predictions for level repulsion in random matrix theory, and indicates that in the symplectic setting there would be $\alpha = 4$ contributing to stronger repulsion. 

\section{Spectra and wavefunctions for the defining representation}
In the most general case (\ref{hamilt}) has a form of a matrix acting on a multicomponent wavefunction, and the dimension of this matrix is equal to the dimension of the $\hat{L}_{ij}$ algebra representation.  Serious difficulties arise when the $\hat{L}_{ij}^2$ operators do not commute thus cannot be simultaneously diagonalized, and the degrees of freedom do not separate. This is why we will focus on the case of the defining representations of $\mathfrak{so}(N)$ and $\mathfrak{su}(N)$, for which the diagonalization of (\ref{hamilt}) with respect to the internal degree of freedom is possible, namely the $N$-dimensional defining representations. In this case:
\begin{equation*}
\left(\hat{L}_{ij}\right)_{ab}= \begin{cases}    \frac{i\hbar}{2}(\delta_{ia}\delta_{jb}-\delta_{ja}\delta_{ib}),\hspace{5mm} i<j, \\
\frac{\hbar}{2}(\delta_{ia}\delta_{jb}+\delta_{ja}\delta_{ib}),\hspace{5mm} i>j.\end{cases}
\end{equation*}
The squared operators are diagonal and (\ref{hamilt})  acts diagonally on the $N$ components of the wavefunction (we shall drop the prime and the subscript for simplicity):
\begin{equation}
\hat{H}\hat{\Psi}=(\hat{H}_1\psi_1,....,\hat{H}_N\psi_N)=E(\psi_1,...,\psi_N)=E\hat{\Psi}.
\end{equation}
The components of the Hamiltonian (together with the harmonic term taken into account in (\ref{harmCMH})) have the following form:
\begin{equation}\label{hamiltI}
\hat{H}_I =\sum_{i=1}^N \left(-\frac{\hbar^2}{2m} \frac{\partial^2}{\partial^2 D_i}+\frac{m\omega^2}{2}D_i^2\right)+\frac{\alpha\hbar^2}{4m}\left(\sum_{i\neq I}\frac{1}{(D_I-D_i)^2}+\sum_{i<j}\frac{\alpha-2}{(D_i-D_j)^2}\right)
\end{equation}

The first sum of interacting terms distinguishes the $\hat{H}_I$ Hamiltonians for different values of $I$: it represents the repulsion between the $I^{th}$ particle and all the others. The second one, which would not be present if we simply promoted the classical $L_{ij}$ variables to quantum operators, represents attraction among all pairs of particles in the orthogonal setting, where $\alpha=1$ and vanishes in the unitary setting with $\alpha=2$. 
\subsection{Separation of variables for arbitrary $N$} 
The next step is to solve the stationary Schr\"odinger equation:
\begin{equation}
\hat{H}_I\psi_I(D_1,...,D_N)=E\psi_I(D_1,...,D_N)
\end{equation}
Similarly as in \cite{calogero69a}, two variables separate, that is the center of mass $R=\frac{1}{N}\sum D_i$, and the classical moment of inertia of the system over mass, that is $I=mr^2$, where
\begin{equation}
r=\sqrt{\frac{1}{N}\sum_{i<j}(D_i-D_j)^2}.
\end{equation}  
The other $N-2$ variables are the angles on the sphere $\mathbb{S}^{N-2}$. The Hamiltonian (\ref{hamiltI}) expressed in these variables:
\begin{eqnarray}
\hat{H}_I&=&\left(-\frac{\hbar^2}{2m}\frac{1}{N}\frac{\partial^2}{\partial R^2}+\frac{m\omega^2 N}{2}R^2\right)+\\
& &+\left[-\frac{\hbar^2}{2m}\left(\frac{\partial^2}{\partial r^2}+\frac{N-2}{r}\frac{\partial}{\partial r}\right)+\frac{m\omega^2 }{2}r^2\right]+\\
& &-\frac{\hbar^2}{2m}\frac{1}{r^2}\left(\Delta_{S^{N-2}}-\frac{\alpha}{2}f_{\alpha,I}(\bar{\varphi})\right),
\end{eqnarray}
where 
\begin{equation}
f_{\alpha,I}(\bar{\varphi})=\left[\sum_{k\neq I}\left(\frac{D_I-D_k}{r}\right)^{-2}+(\alpha-2)\sum_{i<j}\left(\frac{D_i-D_j}{r}\right)^{-2}\right](\bar{\varphi})
\end{equation}
is a function of the angles. The details of this change of variables are shown in the appendix \ref{App:coord}. Postulating the wavefunction in the product form $\psi_I(R,r,\bar{\varphi})=\mathcal{R}(R)\rho(r)\Phi_I(\bar{\varphi})$, we obtain the following equations:
\begin{eqnarray}
\left(-\frac{\hbar^2}{2m}\frac{1}{N}\frac{\partial^2}{\partial R^2}+\frac{m\omega^2 N}{2}R^2\right)\mathcal{R}(R)&=& E_R\mathcal{R}(R) \label{CMass}\\
\left[-\frac{\hbar^2}{2m}\left(\frac{\partial^2}{\partial r^2}+\frac{N-2}{r}\frac{\partial}{\partial r}-\frac{b^2}{r^2}\right)+\frac{m\omega^2 }{2}r^2\right]\rho(r)&=& E_r\rho(r),\label{MInertia} \\
\left(-\Delta_{S^{N-2}}+\frac{\alpha}{2}f_{\alpha,I}(\bar{\varphi})\right)\Phi_I(\bar{\varphi})&=& b^2\Phi_I(\bar{\varphi}),\label{angular}
\end{eqnarray}
and the total energy $E=E_R+E_r$. The solutions of (\ref{CMass}) and (\ref{MInertia}) are well known:
\begin{itemize}
	\item in the free case, $\omega = 0$:
	\begin{eqnarray}
	\mathcal{R}(R)&=& e^{iKR},\hspace{5mm} E_R=\frac{\hbar^2 K^2}{2m} \\
	\rho(r)&=& r^{\frac{3-N}{2}}J_a(kr),\hspace{5mm} a=\sqrt{b^2+\left(\frac{N-3}{2}\right)^2},\hspace{5mm} E_r=\frac{\hbar^2 k^2}{2m},\\
	\end{eqnarray}
	where $J_a(kr)$ are the Bessel functions.
	\item in the harmonic case, where $l=\sqrt{\frac{\hbar}{m\omega}}$:
	\begin{eqnarray}
	\mathcal{R}_n(R)&=& e^{-\frac{NR^2}{2l^2}}H^{(n)}\left(\frac{\sqrt{N}R}{l}\right),\hspace{5mm} E_R=\hbar\omega\left(n+\frac{1}{2}\right) \\
	\rho_{\nu}(r)&=& r^{a-\frac{N-3}{2}} e^{-\frac{r^2}{2l^2}}L_{\nu}^{a}\left(\frac{r^2}{l^2}\right)
	,\hspace{5mm}E_r =\hbar\omega\left(2\nu+a+1\right)
	\end{eqnarray}
	The set of solutions is of-course discrete with $n,\nu =0,1,2...$,  functions $H^{(n)}$ are the Hermite polynomials and $L^a_{\nu}$ are the associated Laguerre polynomials.	
\end{itemize}
Importantly, the dependence on the $\alpha=1,2$ is absent outside the angular equation, which on the other hand does not depend on $\omega$.
\subsection{Complete solution for $N=3$}
The eigenvalue $b^2$ of the angular equation (\ref{angular}) appears in the above solutions through the square root $a=\sqrt{b^2+\left(\frac{N-3}{2}\right)^2}$,  in particular for $N=3$ we have $a=b$. This means that the $r$-dependent component in the free case will be the Bessel function $J_b(kr)$ (with no influence of $b$ on the spectrum), while in the harmonic case $\rho_{\nu}(r)=r^b L^b_{\nu}(\frac{r^2}{l^2})$ and the energy contains an $\hbar\omega b$ term. Moreover, in the three body case (\ref{angular}) simplifies to an ordinary differential equation on a circle and we can find its complete solution. After changing the sign and/or shifting the angular variable defined in the appendix \ref{App:coord} we obtain the following angular equations:
\begin{eqnarray}
\left(-\frac{d^2}{d\varphi^2}-\frac{1}{4\sin^2\varphi}\right)\Phi_{O}(\varphi)&=& b^2_O\Phi_{O}(\varphi) \\
\left(-\frac{d^2}{d\varphi^2}+\frac{1}{4\sin^2\varphi}+\frac{1}{4\sin^2(\varphi-2\pi/3)}\right)\Phi_{U}(\varphi)&=& b^2_U\Phi_{U}(\varphi). 
\end{eqnarray}
The index $I$ is dropped, since it distinguishes only a single repulsive particle, and the different $\Phi_I$ functions are related by shifting one of them by $\pm 2\pi/3$. In the first, orthogonal case the potential is purely attractive, coming from the attracting pair not containing the repulsive particle. 
\subsubsection{The orthogonal case}
We postulate the wavefunction in a form of a Fourier series: $\Phi_O(\phi)=\sum_{n=1}^\infty a_n\sin(n\varphi)$, where $(a_1,a_2,...)\in l^2$, $\sum |a_n|^2 =1$. The symmetry of the $(\sin\varphi)^{-2}$ potential leads to a separation of the odd and even components in the eigenequation:
\begin{eqnarray}
a_{2p-1}\left[(2p-1)^2-p+\frac{1}{2}-b^2\right]-\sum_{q\neq p } a_{2q-1}\left(\min(p,q)-\frac{1}{2}\right)&=& 0 \\
a_{2p}[(2p)^2-p-b^2]-\sum_{q\neq p} a_{2q}\min(p,q)&=& 0
\end{eqnarray}
The infinite-dimensional problem is approximated by a large (dim=100), but finite matrix. The square-roots of the eigenvalues are almost evenly spaced $b_l=l+\delta b_l$ and the eigenfunctions are dominated by the single component: $\psi_l=a_{l,l}\sin(l\varphi)+\delta\psi_l$. In \figurename\ref{fig:orthogonal} the eigenfunctions are presented together with the attractive potential.  The table \tablename\ref{tab:orthogonal} shows the corrections $\delta b$ and $|\delta \psi|$ for the odd and even subspace. Note that the square $|\delta \psi_l|^2$ is the probability of measuring any other wavefunction than $\sin(l\varphi)$ in the $l^{th}$ eigenstate. 
\begin{table}
\begin{minipage}{\textwidth}
		\begin{minipage}[c]{0.48\textwidth}
			\centering
			\vfill
			\includegraphics[width=8cm]{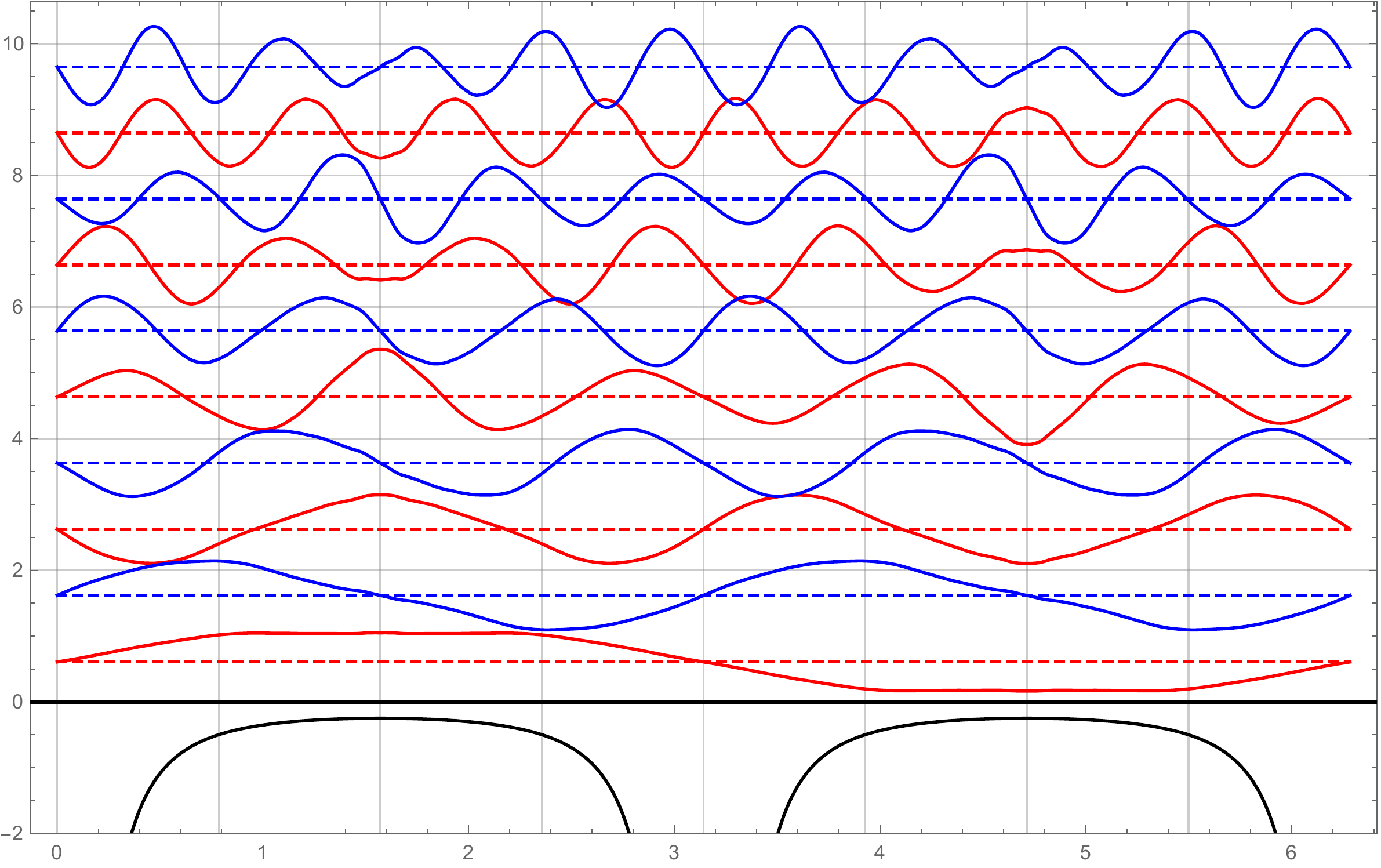}
			\captionof{figure}{The angular eigenfunctions $\psi_l(\varphi)$ and contributions to the energy $b_l$ in the orthogonal case, together with the attractive potential.}\label{fig:orthogonal}
		\end{minipage}
		\hfill
		\begin{minipage}[c]{0.48\textwidth}
			\centering
			\vfill
\begin{tabular}{|c|c|c|c|c|}
	\hline
$p$	& $\delta b_{2p-1}$ & $|\delta\psi_{2p-1}|$ & $\delta b_{2p}$ & $|\delta\psi_{2p}|$  \\ 
	\hline
1	& -0.39 & 0.12 & -0.38 & 0.15 \\ 
2	& -0.38 & 0.20 & -0.37 & 0.22 \\ 
3	& -0.37 & 0.24 & -0.36 & 0.25 \\ 
4	& -0.36 & 0.25 & -0.36 & 0.26 \\ 
5	& -0.36 & 0.26 & -0.35 & 0.26 \\ 
6	& -0.35 & 0.27 & -0.35 & 0.27 \\ 
7	& -0.35 & 0.27 & -0.35 & 0.27\\ 
8	& -0.34 & 0.27 & -0.34 & 0.27\\ 
9	& -0.34 & 0.27 & -0.34 & 0.27\\ 
10	& -0.34 & 0.27 & -0.34 & 0.27\\
	\hline 
\end{tabular}
\captionof{table}{The differences $\delta b_l=b_l-l$ and the norms of $|\delta \psi_l|$, where $\delta \psi_l=\psi_l-a_{l,l}\sin(l\varphi)$.} \label{tab:orthogonal}
\end{minipage}
\end{minipage}
\end{table}
There is also a class of solutions with half-integer values of $b_l=l+\frac{1}{2}$:
\begin{equation}
\psi_{l+\frac{1}{2}}(\varphi)=\sqrt{|\sin\varphi|}\left(A\cdot P_l(\cos\varphi)+B\cdot Q_l(\cos\varphi)\right),
\end{equation}
where $P_l$ and $Q_l$ are the Legendre functions of first and second kind, but they turn out to have a discontinuous probability current at the points, where the potential has a singularity.
\subsubsection{The unitary case}
In the unitary case the potential is repulsive and singular at $\varphi=0,2\pi/3,\pi,5\pi/3$, so we may consider four orthogonal subspaces of wavefunctions: low energy subspaces in the $[0,2\Pi/3]$ and $[\pi,5\pi/3]$ segments and high energy wavefunctions on the $[2\pi/3,\pi]$ and $[5\pi/3,2\pi]$. The longer, low energy segments correspond to both non-interacting particles placed on one side of the repulsive one, and the short, high-energy segments correspond to the repulsive particle placed between the non-interacting ones. The shift $\varphi\rightarrow \varphi+\pi$ corresponds to the exchange of non-interacting particles.

The low-energy solutions on the $[0,2\pi/3]$ segment can be expanded in the following Fourier series: $\Phi_U^L(\varphi)=\sum_{n=1}^{\infty}a_{n,L}\sin\left(\frac{3}{2}n\varphi\right)$, where again $(a_1,a_2,...)_L\in l^2$, $\sum_{n=1}^{\infty}|a_{n,L}|^2=1$. The eigenproblem in the $l^2$ space is the following:
\begin{equation}
a_{n,L}\left(\left(\frac{3n}{2}\right)^2-b^2\right)+\frac{3}{4\pi}\sum_{m=1}^{\infty}a_{m,L} I^L_{m,n}=0,
\end{equation}
where 
\begin{equation}
I^L_{m,n}=(1+(-1)^{m+n})\int_{0}^{2\pi/3}\frac{\sin(3n\varphi/2)\sin(3m\varphi/2)}{\sin^2\varphi}d\varphi
\end{equation}
Once again the off-diagonal values vanish for $m$ and $n$ of different parity, thus odd and even subspaces separate. Let us call the diagonal integrals $I^L_{n}$ as a shorthand. The values of the above integrals are the following:
\begin{eqnarray}
I^L_n&=&\frac{3\pi n}{2}+\frac{\sqrt{3}n}{2}\left(\psi_0\left(\frac{n}{2}+\frac{2}{3}\right)-\psi_0\left(\frac{n}{2}+\frac{1}{3}\right)\right)=\frac{3\pi n}{2}+\frac{\sqrt{3}}{3}-|\delta_n| \\
I^L_{n,m}&=& I^L_{(n+m)/2}-I^L_{|n-m|/2}=\frac{3\pi}{2}\min(n,m)-|\delta_{(n+m)/2}|+|\delta_{|n-m|/2}|,
\end{eqnarray}
where $\psi_0$ is the Digamma function, and $|\delta_n|$ decreases with $n$ as $\frac{8\sqrt{3}}{81n^2}+\mathcal{O}(n^{-4})$, and $|\delta_1|\approx 0.054$. Neglecting these small corrections, we have:
\begin{equation}
a_{n,L}\left(\left(\frac{3n}{2}\right)^2+\frac{9}{8}n+\frac{\sqrt{3}}{4\pi}\right)+\frac{9}{8}\sum_{m\neq n}a_{m,L}\min(n,m)=0 \label{unitaryLOW}
\end{equation}
where both $n$ and $m$ have the same parity.

The Fourier expansion of the high-energy solutions on the $[2\pi/3,\pi]$ segment has the following form: $\Phi_U^H(\varphi)=\sum_{n=1}^{\infty}a_{n,H}\sin\left(3n\varphi\right)$, where again $(a_1,a_2,...)_H\in l^2$, $\sum_{n=1}^{\infty}|a_{n,H}|^2=1$. The eigenproblem in the $l^2$ space is the following:  
\begin{equation}
a_{n,H}\left(\left(3n\right)^2-b^2\right)+\frac{3}{2\pi}\sum_{m=1}^{\infty}a_{m,H} I^H_{m,n}=0,
\end{equation}
where 
\begin{equation}
I^H_{m,n}=(1+(-1)^{m+n})\int_{0}^{\pi/3}\frac{\sin(3n\varphi)\sin(3m\varphi)}{\sin^2\varphi}d\varphi
\end{equation}
The diagonal integrals $I^H_{n,n}$ called $I^H_n$ as a shorthand, thus the off-diagonal ones as well, can be expressed by the low energy integrals:
\begin{eqnarray}
I^H_n&=& 6n\pi-I^L_{2n}=3n\pi-\frac{\sqrt{3}}{3}+|\delta_{2n}| \\
I^H_{n,m} &=& I^H_{(n+m)/2}-I^H_{|n-m|/2}=3\pi\min(n,m)+|\delta_{n+m}|-|\delta_{|n-m|}.
\end{eqnarray}
After neglecting the small corrections contained in the $\delta$ terms, we obtain the final eigenproblem, again :
\begin{equation}
a_{n,H}\left((3n)^2+\frac{9}{2}n-\frac{\sqrt{3}}{2\pi}\right)+\frac{9}{2}\sum_{m\neq n}a_{m,L}\min(n,m)=0 \label{unitaryHIGH}.
\end{equation}
The solutions of (\ref{unitaryLOW}) and (\ref{unitaryHIGH}) on the segment $[0,\pi]$ are presented in \figurename\ref{fig::unitary}. The shifts $\delta b$ in the energy contributions $b^L_l=\frac{3l}{2}+\delta b^L_l$ and $b^H_l=3l+\delta b^L_l$ and the norms of $\delta\psi^{L,H}_l$, where $\psi^L_l=a_{l,l}\sin(3l\varphi/2)+\delta\psi^L_l$ and $\psi^H_l=a_{l,l}\sin(3l\varphi)+\delta\psi^H_l$ are presented in \tablename\ref{tab::unitary}. We note that  (\ref{unitaryLOW}) and (\ref{unitaryHIGH}) differ only by a factor and a shift, so they have the same eigenvectors, therefore $|\delta\psi^L|=|\delta\psi^H|=|\delta\psi|$.

\begin{table}
	\begin{minipage}{\textwidth}
		\begin{minipage}[c]{0.48\textwidth}
			\centering
			\vfill
			\includegraphics[width=8cm]{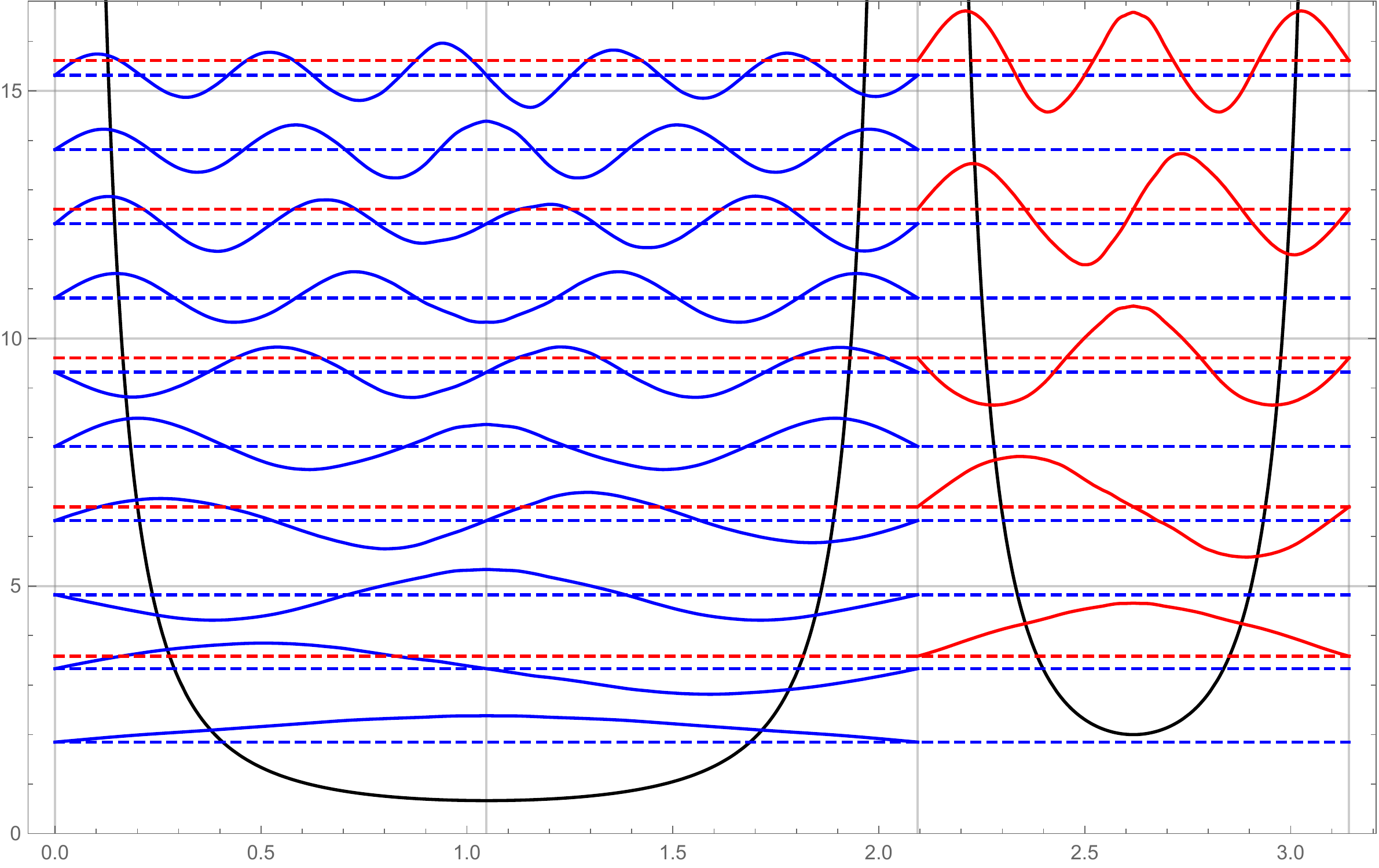}
			\captionof{figure}{The angular eigenfunctions $\psi_l(\varphi)$ and contributions to the energy $b_l$ in the unitary case, together with the repulsive potential.}\label{fig::unitary}
		\end{minipage}
		\hfill
		\begin{minipage}[c]{0.48\textwidth}
			\centering
			\vfill
			\begin{tabular}{|c|c|c|c|}
				\hline
				$l$	& $\delta b^L_l$ & $\delta b^H_l$ & $|\delta\psi|$ \\ 
				\hline
				1	& 0.35 & 0.58 & 0.05  \\ 
				2	& 0.33 & 0.60 & 0.07  \\ 
				3	& 0.33 & 0.61 & 0.10  \\ 
				4	& 0.32 & 0.61 & 0.11  \\ 
				5	& 0.32 & 0.61 & 0.12  \\ 
				6	& 0.32 & 0.61 & 0.13  \\ 
				7	& 0.32 & 0.62 & 0.14 \\ 
				8	& 0.32 & 0.62 & 0.14 \\ 
				9	& 0.32 & 0.62 & 0.14 \\ 
				10	& 0.32 & 0.62 & 0.15 \\
				\hline 
			\end{tabular}
			\captionof{table}{The differences $\delta b^L_l=b^L_l-3l/2$ and $\delta b^H_l=b^H_l-3l$ and the norms of $|\delta \psi^L|=|\delta \psi^H|$, where $\delta\psi=\sqrt{1-a_{l,l}^2}$.} \label{tab::unitary}
		\end{minipage}
	\end{minipage}
\end{table}

\section{Conclusions and outlook}
We have derived the quantum Calogero-Moser Hamiltonian (\ref{hamilt}) by applying the reduction procedure directly based on the classical one. The result differs from what we would get by simply upgrading the $L_{ij}$ classical variables to operators with the commutation relations given by the Poisson brackets. The difference lies in the $\sum_{i<j}\frac{\hbar^2\alpha(\alpha-2)}{4m(D_i-D_j)^2}$ terms with $\alpha=1$ in the orthogonal and $\alpha=2$ in the unitary setting. In the unitary setting these terms of course vanish, but in the orthogonal case they give raise to an attractive potential. This leads to a much richer set of eigenstates than the simple canonical quantization, even in the case as simple as $N=2$. Moreover, the weaker repulsion is a trait of the classical orthogonal CM system as compared to the unitary one. The reduction of the free quantum system restores this difference on the quantum level, in contrast to the straightforward canonical quantization, which makes us loose this distinction. This yields the reduction method hugely advantageous.

 For more than two particles the non-commutativity of the $\hat{L}_{ij}$ operators starts to play a role, and particular representations of the algebra are necessary to separate the degrees of freedom. In the defining representation,in both the orthogonal and unitary case, the $\hat{L}^2_{ij}$ operators turn out to have a common diagonalizing basis, and we have considered the Hamiltonian given by this representation. We solved it completely for the smallest nontrivial case of $N=3$, and shown the significant influence of the $\alpha$-dependent terms on both the spectra and the wavefunctions. 

The presented results motivate further research in the following directions: 1. The search of a basis of functions on the $N-2$ dimensional sphere, in order to solve the eigenproblem (\ref{angular}) for a general $N$, 2. The search of other representations of $\mathfrak{so}(N)$ and $\mathfrak{su}(N)$ algebra, for which the spatial and internal degrees of freedom in (\ref{hamilt}) will separate, 3. Extension of the reduction procedure from real ($\alpha=1=2^0$) and complex ($\alpha=2=2^1$) Hermitian matrices to hypercomplex numbers $\alpha=2^{\nu}$, especially the symplectic $\alpha=4$ (corresponding to quaternions), for which the classical results are known. 
\appendix
\section{Details of proofs in \ref{SEC:QUANTRED}}\label{App:B}
\subsection{Derivation of the formulae (\ref{Jac})-(\ref{g})}\label{derivDaH}
Let us start with the entries of the Jacobian matrix (\ref{Jac}):
\begin{eqnarray*}
	\frac{\partial X_{ij}}{\partial D_k}&=&\frac{\partial}{\partial D_k}\left(U^{\dagger}DU\right)_{ij}=\frac{\partial}{\partial D_k}\left(\sum_{m=1}^{N}U^{\dagger}_{im}D_mU_{mj}\right)=U^*_{ki}U_{kj}\\
	\frac{\partial X_{ij}}{\partial a_l}&=&\frac{\partial}{\partial a_l}\left(U^{\dagger}(\bar{a})DU(\bar{a})\right)_{ij}=\left(\partial_{a_l}U^{\dagger}DU+U^{\dagger}D\partial_{a_l}U\right)_{ij}=\left(U^{\dagger}D\partial_{a_l}U-U^{\dagger}(\partial_{a_l}U)U^{\dagger}DU\right)_{ij}=\\
	&=&\left[U^{\dagger}\left(D(\partial_{a_l}U)U^{\dagger}-(\partial_{a_l}U)U^{\dagger}D\right)U\right]_{ij}=\left(U^{\dagger}\left[D,(\partial_{a_l}U)U^{\dagger}\right]U\right)_{ij}=\left(U^{\dagger}\Omega_lU\right)_{ij}
\end{eqnarray*}
Next, we derive the expression for $(\partial_{a_l}U)U^{\dagger}$ stated in (\ref{omegas}):
\begin{eqnarray*}
	(\partial_{a_l}U)U^{\dagger}&=&-U\partial_{a_l}U^{\dagger}=-e^{\bar{a}\cdot\bar{\tau}}\partial_{a_l}e^{-\bar{a}\cdot\bar{\tau}}=-e^{\bar{a}\cdot\bar{\tau}}e^{-\bar{a}\cdot\bar{\tau}}\left(\frac{\mathbb{1}-e^{-(-ad_{\bar{a}\cdot\bar{\tau}})}}{-ad_{\bar{a}\cdot\bar{\tau}}}\right)\partial_{a_l}(-\bar{a}\cdot\bar{\tau})=\\
	&=&\left(\frac{e^{ad_{\bar{a}\cdot\bar{\tau}}}-\mathbb{1}}{ad_{\bar{a}\cdot\bar{\tau}}}\right)\tau_l=\sum_{n=0}^{\infty}\frac{(ad_{\bar{a}\cdot\bar{\tau}})^n}{(n+1)!}\tau_l=\\
	&=&\sum_{n=0}^{\infty}\frac{1}{(n+1)!}\underbrace{\left[\bar{a}\cdot\bar{\tau},\left[\bar{a}\cdot\bar{\tau},...,\left[\bar{a}\cdot\bar{\tau},\tau_l\right]...\right]\right]}_n=\\
	&=&\sum_{n=0}^{\infty}\frac{1}{(n+1)!}a_{k_n}f_{k_n,m_{n-1}}^{m_n}....a_{k_2}f_{k_2,m_1}^{m_2}a_{k_1}f_{k_1,l}^{m_1}\tau_{m_n}=\\
	&=&\sum_{n=0}^{\infty}\frac{(a_kf_k)^n_{lm}}{(n+1)!}\tau_{m}=\sum_{n=0}^{\infty}\frac{(A^n)_{lm}}{(n+1)!}\tau_m
	=u(A)_{lm}\tau_m \in \mathfrak{g}\\
	(\Omega_l)_{ij}&=& u(A)_{lm}\left[D,\tau_m\right]_{ij}=u(A)_{lm}(D_i-D_j)(\tau_m)_{ij},
\end{eqnarray*}
where we make use of a general formula for a smooth matrix valued function $X(t)$:
\begin{equation}
\frac{de^{X(t)}}{dt}=e^{X(t)}\left[\frac{\mathbb{1}-e^{-ad_{X(t)}}}{ad_{X(t)}}\left(\frac{dX(t)}{dt}\right)\right].
\end{equation}
Importantly, since we use anti-Hermitian generators, the structure constants $f_{ij}^k\in\mathbb{R}$, thus $u(A)$ is a real matrix in both unitary and orthogonal case.
Now we calculate the entries of the metric tensor (\ref{gDa}):
\begin{eqnarray}
g_{D_k,D_m}&=&\sum_{i,j}\frac{\partial X_{ij}}{\partial D_k}\frac{\partial X_{ji}}{\partial D_m}=\sum_{i,j}U^*_{ki}U_{kj}U^*_{mj}U_{mi}=\delta_{km}\\
g_{D_k,a_l}&=&\sum_{i,j}\frac{\partial X_{ij}}{\partial D_k}\frac{\partial X_{ji}}{\partial a_l}=\sum_{ij}U^*_{ki}U_{kj}\left(U^{\dagger}\Omega_lU\right)_{ji}=\left(\Omega_l\right)_{kk}=0\\
g_{a_l,a_m}&=&\sum_{i,j}\frac{\partial X_{ij}}{\partial a_l}\frac{\partial X_{ji}}{\partial a_m}=\sum_{i,j}\left(U^{\dagger}\Omega_lU\right)_{ij}\left(U^{\dagger}\Omega_mU\right)_{ji}=\tr(\Omega_l\Omega_m)
\end{eqnarray}
We notice that the nontrivial block $g_{lm}=Tr(\Omega_l\Omega_m)$ of the metric tensor is in fact a Gram matrix of $\Omega_{1,2,...,d}$ treated as vectors in the vector space of Hermitian matrices. This means that the determinant can be expressed with the exterior product of the vectors:
\begin{equation}
\mathrm{det} g = ||\Omega_1\wedge\Omega_2\wedge...\wedge\Omega_d||^2.
\end{equation}
To continue with the proofs of (\ref{g})-(\ref{g1}), we need to switch the indices to ordered pair indices, as in the main text, $(ij),(pq),(rs)\in I$, where $I=\lb(pq): 1\leq p < q \leq N\rb$ in the orthogonal case and $I=\lb(pq): 1\leq p \neq q \leq N\rb$ in the unitary case, and express the $\Omega$ matrices through anti-Hermitian generators:
\begin{equation}
\tau_{ij}=\left\{ \begin{array}{c}
\frac{1}{2}\left(|i\rangle\langle j|-|j\rangle\langle i|\right),\hspace{5mm} i<j \\
\frac{i}{2}\left(|i\rangle\langle j|+|j\rangle\langle i|\right),\hspace{5mm} i>j
\end{array}\right. ,\hspace{5mm} Tr(\tau_{ij}\tau_{kl})=-\frac{1}{2}\delta_{(ij)(kl)}.\label{tauij}
\end{equation}
Let us note that these generators obey the commutation relations given by (\ref{pbo}) and (\ref{pbu}). The expressions for $\Omega_{ij}$ and $g_{(ij)(kl)}$ are as follows:
\begin{eqnarray*}
	\Omega_{ij}&=&\sum_{(pq)\in I}u_{(ij)(pq)}\left[D,\tau_{pq}\right]=\sum_{(pq)\in I}u_{(ij)(pq)}(D_p-D_q)(-i\tau_{pq})\\
	g_{(ij)(kl)}&=& Tr(\Omega_{ij}\Omega_{kl})=\sum_{(pq)\in I}\sum_{(rs)\in I}u_{(ij)(pq)}(D_p-D_q)u_{(kl)(rs)}(D_r-D_s)Tr(-\tau_{pq}\tau_{rs})=\\
	&=& \frac{1}{2}\sum_{(pq)\in I}\sum_{(rs)\in I}u_{(ij)(pq)}(D_p-D_q)u_{(kl)(rs)}(D_r-D_s)\delta_{pr}\delta_{qs}=\\
	&=& \frac{1}{2}\sum_{(pq)\in I}u_{(ij)(pq)}u_{(kl)(pq)}(D_p-D_q)^2=\frac{1}{2}\sum_{(pq)\in I}u_{(ij)(pq)}(D_p-D_q)^2(u^T)_{(pq)(kl)}=\\
	&=& \frac{1}{2}(u \bold{D}^2 u^T)_{(ij)(kl)}.
\end{eqnarray*}
The factorised form of $g$ given by (\ref{g}) leads automatically to (\ref{detg}),(\ref{g1}) and the expressions for $\Delta_a$ and $\hat{\Lambda}_{ij}$. 

\subsection{Derivation of (\ref{SIMD})}\label{derSIMD}
The action of the similarity transformation (\ref{SIMD}) can be demonstrated with the use of a $\mathcal{C}^2$ class test function $f=f(D_1,D_2,...,D_N)$, and the fact that:
\begin{equation}
\frac{\partial\mathcal{D}}{\partial D_i}=\frac{\partial}{\partial D_i}\left(\prod_{k<l}(D_k-D_l)^{\alpha}\right)=\alpha\mathcal{D}\left(\sum_{k\neq i}\frac{1}{D_i-D_k}\right).
\end{equation}
The calculations are as follows:
{\small 
	\begin{eqnarray*}
		\sqrt{\mathcal{D}}\Delta_D\left(\frac{f}{\sqrt{\mathcal{D}}}\right)&=&\frac{1}{\sqrt{\mathcal{D}}}\sum_{i=1}^N\frac{\partial}{\partial D_i}\left[\mathcal{D}\frac{\partial}{\partial D_i}\left(\frac{f}{\sqrt{\mathcal{D}}}\right)\right]=
		\sum_{i=1}^N\left\lb \frac{\partial^2 f}{\partial D_i^2}+a_i(D)\left(\frac{\partial f}{\partial D_i}\right)\right\rb+ b(D)f \\
		\sqrt{\mathcal{D}}a_i(D)&=&\frac{\partial\sqrt{\mathcal{D}}}{\partial D_i}+\mathcal{D}\frac{\partial}{\partial D_i}\left(\frac{1}{\sqrt{\mathcal{D}}}\right)=\frac{1}{2\sqrt{\mathcal{D}}}\frac{\partial\mathcal{D}}{\partial D_i}-\mathcal{D}\frac{1}{2\sqrt{\mathcal{D}}^3}\frac{\partial\mathcal{D}}{\partial D_i}=0 \\
		b(D)&=&\sum_{i=1}^N\frac{1}{\sqrt{\mathcal{D}}}\frac{\partial}{\partial D_i}\left[\mathcal{D}\frac{\partial}{\partial D_i}\left(\frac{1}{\sqrt{\mathcal{D}}}\right)\right]=\sum_{i=1}^N\frac{1}{\sqrt{\mathcal{D}}}\frac{\partial}{\partial D_i}\left[-\mathcal{D}\left(\frac{1}{2\sqrt{\mathcal{D}}^3}\frac{\partial\mathcal{D}}{\partial D_i}\right)\right]=\\
		&= &-\frac{\alpha}{2\sqrt{\mathcal{D}}}\sum_{i=1}^N\frac{\partial}{\partial D_i}\left(\sqrt{\mathcal{D}}\sum_{k\neq i}\frac{1}{D_i-D_k}\right)=\\
		&=&-\frac{\alpha}{2\sqrt{\mathcal{D}}}\sum_{i=1}^N\left[\frac{1}{2\sqrt{\mathcal{D}}}\frac{\partial\mathcal{D}}{\partial D_i}\sum_{k\neq i}\frac{1}{D_i-D_k}-\sqrt{\mathcal{D}}\sum_{k\neq i}\frac{1}{(D_i-D_k)^2}\right]=\nonumber\\
		&=&\frac{\alpha}{2}\sum_{1\leq i\neq j\leq N}\frac{1}{(D_i-D_j)^2}-\frac{\alpha^2}{4}\left[\sum_{1\leq i\neq j\leq N}\frac{1}{(D_i-D_j)}\right]^2=\nonumber\\
		&=&\frac{\alpha}{2}\left(1-\frac{\alpha}{2}\right)\sum_{1\leq i\neq j\leq N}\frac{1}{(D_i-D_j)^2}-\frac{\alpha^2}{4}\sum_{i<j<k}\left[\frac{D_k-D_j+D_i-D_k+D_j-D_i}{(D_i-D_j)(D_j-D_k)(D_k-D_i)}\right]\nonumber\\
		&=&\frac{\alpha\left(2-\alpha\right)}{2}\sum_{1\leq i\leq j\leq N}\frac{1}{(D_i-D_j)^2}=\frac{1}{2}\sum_{(ij)}\frac{2-\alpha}{(D_i-D_j)^2}\nonumber
	\end{eqnarray*}}
	
	\subsection{Derivation of (\ref{LAMBDA2}) and (\ref{LIJKL})}\label{derL}
	Let $M$ be a $d\times d$ matrix of real functions: $M_{ij}: \mathbb{R}^d\rightarrow \mathbb{R}$. At every point $(x_1,..,x_d)\in\mathbb{R}^d$ where $M$ is invertible, we may say that:
	\begin{equation*}
	M^{-1}=\frac{C(M)^T}{\mathrm{det}M}
	\end{equation*}
	where $C(M)$ is the matrix of cofactors of $M$. This means that:
	\begin{eqnarray*}
		\mathrm{det}M\delta_{ij}&=&\sum_{l=1}^d M_{il}C(M)_{jl} \\
		\partial_i (\mathrm{det}M) &=&\sum_{j=1}^d \partial_j(\mathrm{det}M)\delta_{ij}= \sum_{j,l=1}^d\partial_j( M_{il}C(M)_{jl}) \\
		&=& \sum_{j,l=1}^d \partial_j(M_{il})C(M)_{jl}+M_{il}\partial_j(C(M)_{jl}).
	\end{eqnarray*}
	On the other hand:
	\begin{equation}
	\partial_i(\mathrm{det}M)=\sum_{j,l=1}^d \frac{\partial \mathrm{det} M}{\partial M_{jl}}\frac{\partial M_{jl}}{\partial x_i}=\sum_{j,l=1}^d C(M)_{jl}\partial_i(M_{jl}),
	\end{equation}
	which means that
	\begin{equation}\label{Mij}
	\sum_{l=1}^d M_{il}\sum_{j=1}^d \partial_j(C(M)_{jl})=\sum_{j,l=1}^d C(M)_{jl}(\partial_i M_{jl}-\partial_j M_{il}).
	\end{equation}
	If $M^T$ happens to be a Jacobian matrix of some map $m: \mathbb{R}^d\rightarrow \mathbb{R}^d$, the right-hand side of the above equation vanishes identically:
	\begin{equation}
	\partial_i M_{jl}-\partial_j M_{il}=\partial^2_{ij}m_l-\partial^2_{ji}m_l =0 .
	\end{equation}
	For the left-hand side to vanish, the sum $F_l = \sum_{j=1}^d \partial_j(C(M)_{jl})$ must vanish identically as well. This is so, because wherever $\mathrm{det}M\neq 0$, $M_i=(M_{i1},M_{i2},...,M_{id})$ can be treated as $d$ linearly independent vectors and for the scalar product $(M_i,F)=\sum_{l} M_{il} F_l$ to vanish for all of them $F=0$ identically. This proof can be found for example in \cite{evans}.
	
	Let us now translate the equation (\ref{Mij}) to the language of the $u$ matrix and $a_{ij}$ variables. The indices change in the following way: $j\rightarrow pq$, $l\rightarrow ij$, $i\rightarrow ab$:
	\begin{equation}\label{Fij0}
	(\mathrm{det} u)\sum_{(ij)\in I}u_{(ab)(ij)}F_{ij}=\sum_{(ij),(pq)\in I} C(u)_{(pq)(ij)}\left(\frac{\partial u_{(pq)(ij)}}{\partial a_{ab}}-\frac{\partial u_{(ab)(ij)}}{\partial a_{pq}}\right).
	\end{equation}
	The matrix elements of $u$ can be derived from the definition:
	\begin{equation}
	\frac{\partial U}{\partial a_{pq}}U^{\dagger}=\sum_{(ij)\in I}u_{(pq)(ij)}\tau_{ij}\implies u_{(pq)(ij)}=-2\tr\left[\frac{\partial U}{\partial a_{pq}}U^{\dagger}\tau_{ij}\right]
	\end{equation}
	and its derivatives have the following form:
	\begin{eqnarray}
	\frac{\partial u_{(rs)(mn)}}{\partial a_{pq}}&=&-2\tr\left[\left(\frac{\partial^2 U}{\partial a_{pq}\partial a_{rs}}\right)U^{\dagger}\tau_{mn}\right]-2\tr\left(\frac{\partial U}{\partial a_{rs}}\frac{\partial U^{\dagger}}{\partial a_{pq}}\tau_{mn}\right)= \nonumber\\
	&=&-2\tr\left[\left(\frac{\partial^2 U}{\partial a_{pq}\partial a_{rs}}\right)U^{\dagger}\tau_{mn}\right]-2tr\left(\frac{\partial U}{\partial a_{rs}}U^{\dagger}U\frac{\partial U^{\dagger}}{\partial a_{pq}}\tau_{mn}\right)\nonumber = \\
	&=&-2\tr\left[\left(\frac{\partial^2 U}{\partial a_{pq}\partial a_{rs}}\right)U^{\dagger}\tau_{mn}\right]+2\tr\left(\frac{\partial U}{\partial a_{rs}}U^{\dagger}\frac{\partial U}{\partial a_{pq}}U^{\dagger}\tau_{mn}\right)\nonumber = \\
	&=&-2\tr\left[\left(\frac{\partial^2 U}{\partial a_{pq}\partial a_{rs}}\right)U^{\dagger}\tau_{mn}\right]+\label{usym}\\
	& &+2\sum_{(ab),(cd)\in I}u_{(rs)(ab)}u_{(pq)(cd)}\tr(\tau_{ab}\tau_{cd}\tau_{mn})\label{uasym}
	\end{eqnarray}
	where we again use the definition of $u$ and the fact that $(\partial U)U^{\dagger}=-U\partial U^{\dagger}$. The symmetric parts (\ref{usym}) cancel in the difference of derivatives:
	\begin{eqnarray}
		\frac{\partial u_{(pq)(ij)}}{\partial a_{ab}}-\frac{\partial u_{(ab)(ij)}}{\partial a_{pq}}
		&=&\sum_{(mn),(rs)\in I}2u_{(pq)(mn)}u_{(ab)(rs)}\tr([\tau_{mn},\tau_{rs}]\tau_{ij})=\nonumber\\
		&=&-\sum_{(mn),(rs)\in I}u_{(pq)(mn)}u_{(ab)(rs)}f_{(mn)(rs)}^{(ij)}=\nonumber\\
		&=&(uf^{(ij)}u^T)_{(ab)(pq)},\label{ufiju}
	\end{eqnarray}
	
	This difference does not vanish identically, which means $u^T$ is not a Jacobian matrix of any map, nevertheless we may apply it to the right-hand side of (\ref{Fij0}):
	\begin{eqnarray*}
		\sum_{(ij),(pq)\in I} (uf^{(ij)}u^T)_{(ab)(pq)}C(u)_{(pq)(ij)}&=&\sum_{(ij),(pq)\in I} (uf^{(ij)}u^T)_{(ab)(pq)}(u^T)^{-1}_{(pq)(ij)}\mathrm{det} u =\\
		&=& \mathrm{det} u \sum_{(ij)\in I} (u f^{(ij)})_{(ab)(ij)}=0.
	\end{eqnarray*}
	It turns out to be $0$ due to the fact that the structure constants are antisymmetric in every pair of indices and $f_{(ij)(kl)}^{(ij)}=0$. Using the same arguments as for the general matrix $M$ we conclude that for the left-hand side of (\ref{Fij0}) to vanish, the functions $F_{ij}$ must vanish as well.
	
	The commutator of $\hat{\lambda}$ operators can be calculated in the following way:
	\begin{eqnarray*}
		\left[\hat{\lambda}_{ij},\hat{\lambda}_{kl}\right]&=&\sum_{(pq),(rs)\in I}\left[u^{-1}_{(ij)(pq)}\partial_{pq},u^{-1}_{(kl)(rs)}\partial_{rs}\right]=\sum_{(rs)\in I}\mu_{(ij)(kl)}^{(rs)}\partial_{rs}=\nonumber\\
		&=&\sum_{(mn)}\nu_{(ij)(kl)}^{(mn)}\hat{\lambda}_{mn}\\
		\mu_{(ij)(kl)}^{(rs)}&=&\sum_{(pq)\in I}u^{-1}_{(ij)(pq)}(\partial_{pq}u^{-1}_{(kl)(rs)})-u^{-1}_{(kl)(pq)}(\partial_{pq}u^{-1}_{(ij)(rs)}) \\
		\sum_{(rs)}\mu_{(ij)(kl)}^{(rs)}\partial_{rs}&=&\sum_{(rs)(mn),(uv)} \mu_{(ij)(kl)}^{(rs)}u_{(rs)(mn)}u^{-1}_{(mn)(uv)}\partial_{uv} =\sum_{(mn)}\nu_{(ij)(kl)}^{(mn)}\hat{\lambda}_{mn}\nonumber \\
		\nu_{(ij)(kl)}^{(mn)}&=&\sum_{(rs)\in I} \mu_{(ij)(kl)}^{(rs)}u_{(rs)(mn)}
	\end{eqnarray*}
	where $\partial_{pq}$ stands for $\frac{\partial}{\partial a _{pq}}$ and for the final calculation of $\nu$ we use the formulae (\ref{usym}), (\ref{uasym}) and (\ref{ufiju}):
	\begin{eqnarray*}
		\nu_{(ij)(kl)}^{(mn)}&=&\sum_{(rs)(pq)}u^{-1}_{(ij)(pq)}(\partial_{pq}u^{-1}_{(kl)(rs)})-u^{-1}_{(kl)(pq)}(\partial_{pq}u^{-1}_{(ij)(rs)})u_{(rs)(mn)}=\\
		&=&\sum_{(rs)(pq)}u^{-1}_{(ij)(pq)}\left[\partial_{pq}(u^{-1}_{(kl)(rs)}u_{(rs)(mn)})-u^{-1}_{(kl)(rs)}\partial_{pq}u_{(rs)(mn)}\right]+\\
		& -&\sum_{(rs)(pq)}u^{-1}_{(kl)(pq)}\left[\partial_{pq}(u^{-1}_{(ij)(rs)}u_{(rs)(mn)})-u^{-1}_{(ij)(rs)}\partial_{pq}u_{(rs)(mn)}\right]=\\
		&=&\sum_{(pq),(rs)}\left(u^{-1}_{(kl)(pq)}u^{-1}_{(ij)(rs)}-u^{-1}_{(ij)(pq)}u^{-1}_{(kl)(rs)}\right)\partial_{pq}u_{(rs)(mn)}=\\
		&=&\sum_{(pq),(rs)}u^{-1}_{(kl)(pq)}u^{-1}_{(ij)(rs)}\left(\partial_{pq}u_{(rs)(mn)}-\partial_{rs}u_{(pq)(mn)}\right)=\\
        &=&\sum_{(pq),(rs)} u^{-1}_{(kl)(pq)}u^{-1}_{(ij)(rs)}\left(uf^{(mn)}u^T\right)_{(pq)(rs)}=\\
        &=& \left(u^{-1}u f^{(mn)}u^T(u^T)^{-1}\right)_{(kl)(ij)}=f^{(mn)}_{(kl)(ij)}=-f^{(mn)}_{(ij)(kl)}
	\end{eqnarray*}

\section{Details of the $(R,r,\varphi_j)$ coordinate transformation}\label{App:coord}
 After separating the center of mass coordinate of the $N$-body system, we can describe the relative positions with $N-1$ Jacobi coordinates, which in the case of equal masses are (up to a proportionality constant) equal to:
 \begin{eqnarray*}
 D_1-D_2 &=& \sqrt{2}x_1 \\
 D_1+D_2-2D_3 &=& \sqrt{6}x_2 \\
 .... \\
 D_1+...+D_{N-1}-(N-1)D_N &=&\sqrt{N(N-1)}x_{N-1}.
 \end{eqnarray*}
 Next, since $\sum_{i=1}^{N-1} x_i^2 = r^2$, we can use $r$ and $N-2$ angles to describe each $(x_1,...,x_{N-1})$ as a point on a sphere:
 \begin{equation}
 (x_1,x_2,...,x_{N-2},x_{N-1})=(r\cos\varphi_1,r\sin\varphi_1\cos\varphi_2,...,r\sin\varphi_{1,...,N-3}\cos\varphi_{N-2},r\sin\varphi_{1,...,N-2}),
 \end{equation}
 and these are exactly the coordinates we need. The complete transformation, where $c_i=\cos\varphi_i$ and $s_i=\sin\varphi_i$  :
 \begin{equation}
 \left(\begin{array}{c} D_1 \\ D_2 \\ ... \\ ... \\ ... \\ D_N \end{array}\right) =\left( \begin{array}{c c c c c c} 
 \frac{1}{\sqrt{2}} & \frac{1}{\sqrt{6}}& ... & ... & \frac{1}{\sqrt{N(N-1)}} & \frac{1}{\sqrt{N}} \\
 -\frac{1}{\sqrt{2}} & \frac{1}{\sqrt{6}}& ... & ... & \frac{1}{\sqrt{N(N-1)}} & \frac{1}{\sqrt{N}} \\
 0 & -\frac{2}{\sqrt{6}}& ... & ... & \frac{1}{\sqrt{N(N-1)}} & \frac{1}{\sqrt{N}} \\
 ... & ... & ... & ... & ... & ... \\
 0   & ... & 0 &  -\sqrt{\frac{N-2}{N-1}}  & \frac{1}{\sqrt{N(N-1)}} & \frac{1}{\sqrt{N}} \\
  0   & ... & 0 &  0 & -\sqrt{\frac{N-1}{N}} & \frac{1}{\sqrt{N}} \end{array}
 \right) \left(\begin{array}{c} rc_1 \\ rs_1c_2 \\ ... \\ ... \\ rs_{1,...,N-2} \\ \sqrt{N}R \end{array}\right),
\end{equation}
can be written shortly as $\bar{D}=M(r,\bar{\varphi},R)$, where $M$ is an orthogonal matrix. The metric tensor:
\begin{equation}
g = \left(\begin{array}{c c c c c}
g_{RR} & g_{Rr} & g_{R\varphi_1} & ... & g_{R\varphi_{N-2}} \\
g_{rR} & g_{rr} & g_{r\varphi_1} & ... & g_{r\varphi_{N-2}} \\
g_{\varphi_1 R} &  g_{\varphi_1 r} & ... & ... & ... \\
... & ... & ... & g_{\varphi_i\varphi_j} & ... \\
g_{\varphi_{N-2}R} &  g_{\varphi_{N-2 }r} & ... & ... & ...

 \end{array}\right)= \left(\begin{array}{c c c c c}
 N & 0 & 0 & ... & 0 \\
 0 & 1 & 0 & ... & 0 \\
 0&  0 & ... & ... & ... \\
 ... & ... & ... & r^2 \partial_{\varphi_i}\bar{\varphi}\partial_{\varphi_j}\bar{\varphi} & ... \\
 0 &  0 & ... & ... & ...
 \end{array}\right),
\end{equation}
where $\bar{\varphi}=(c_1,s_1c_2,...,s_{1,2,...,N-2})$, leads to the know form of the Laplacian. There is a simple expression for the harmonic term $\sum_{i=1}^N D_i^2 = NR^2+r^2$, and since for all the differences $D_i-D_j=rf_{ij}(\bar{\varphi})$, the functions $f_{\alpha,I}$ depend only on the angles.

\section*{Acknowledgements}

The authors acknowledge a financial support of the the Polish National Science Centre grant 2017/27/B/ST2/02959.

\bibliographystyle{unsrt}

\begin{thebibliography}{10}
	
	\bibitem{calogero69}
	F.~Calogero.
	\newblock {Solution of a Three‐Body Problem in One Dimension}.
	\newblock {\em Journal of Mathematical Physics}, 10(12):2191--2196, 1969.
	
	\bibitem{calogero69a}
	F.~Calogero.
	\newblock {Ground State of a One‐Dimensional N‐Body System}.
	\newblock {\em Journal of Mathematical Physics}, 10(12):2197--2200, 1969.
	
	\bibitem{calogero71}
	F.~Calogero.
	\newblock {Solution of the One‐Dimensional N‐Body Problems with Quadratic
		and/or Inversely Quadratic Pair Potentials}.
	\newblock {\em Journal of Mathematical Physics}, 12(3):419--436, 1971.
	
	\bibitem{moser75}
	J~Moser.
	\newblock Three integrable {H}amiltonian systems connected with isospectral
	deformations.
	\newblock {\em Advances in Mathematics}, 16(2):197 -- 220, 1975.
	
	\bibitem{kazhdan78}
	David Kazhdan, Bertram Kostant, and Shlomo Sternberg.
	\newblock Hamiltonian group actions and dynamical systems of calogero type.
	\newblock {\em Communications on Pure and Applied Mathematics}, 31(4):481--507,
	1978.
	
	\bibitem{marsden74}
	J.~Marsden and A.~Weinstein.
	\newblock Reduction of symplectic manifolds with symmetry.
	\newblock {\em {Reports on Mathematical Physics}}, 5(1):121--130, 1974.
	
	\bibitem{kowalczykmurynka22}
	K.~Kowalczyk-Murynka and M.~Ku{\'s}.
	\newblock {Matrix and vectorial generalized Calogero{\textendash}Moser models}.
	\newblock {\em Physica D}, 440:133491, November 2022.
	
	\bibitem{sutherland71}
	B.~Sutherland.
	\newblock Exact results for a quantum many-body problem in one dimension.
	\newblock {\em Phys. Rev. A}, 4:2019--2021, 1971.
	
	\bibitem{sutherland72}
	B.~Sutherland.
	\newblock Exact results for a quantum many-body problem in one dimension. ii.
	\newblock {\em Phys. Rev. A}, 5:1372--1376, 1972.
	
	\bibitem{calogero75}
	F.~Calogero, O.~Ragnisco, and C.~Marchioro.
	\newblock Exact solution of the classical and quantal one-dimensional many-body
	problems with the two-body potential $v_a(x)=g^2a^2/\mathrm{sh}^2(ax)$.
	\newblock {\em Lettere al Nuovo Cimento}, 13:383--387, 1975.
	
	\bibitem{olshanetsky81}
	M. A. Olshanetsky and A. M. Perelomov.
	\newblock {Classical integrable finite-dimensional systems related to Lie
		algebras}.
	\newblock {\em Physics Reports}, 71(5):313 -- 400, 1981.
	
	\bibitem{wojciechowski85}
	S.~Wojciechowski.
	\newblock An integrable marriage of the euler equations with the calogero-moser
	system.
	\newblock {\em Physics Letters A}, 111(3):101 -- 103, 1985.
	
	\bibitem{gibbons84}
	J.~Gibbons and T.~Hermsen.
	\newblock {A generalisation of the Calogero-Moser system}.
	\newblock {\em Physica D: Nonlinear Phenomena}, 11(3):337 -- 348, 1984.
	
	\bibitem{pechukas83}
	P.~Pechukas.
	\newblock Distribution of energy eigenvalues in the irregular spectrum.
	\newblock {\em Physical review letters}, 51(11):943, 1983.
	
	\bibitem{yukawa85}
	T~Yukawa.
	\newblock New approach to the statistical properties of energy levels.
	\newblock {\em Physical review letters}, 54(17):1883, 1985.
	
	\bibitem{haake19}
	F.~Haake, S~Gnutzmann, and M~Ku\'s.
	\newblock {\em {Quantum Signatures of Chaos}}.
	\newblock Springer, International Publishing, 2019.
	
	\bibitem{feher08}
	L~Feh{\'e}r and B.~G. Pusztai.
	\newblock Hamiltonian reductions of free particles under polar actions of
	compact {L}ie groups.
	\newblock {\em Theoretical and Mathematical Physics}, 155(1):646--658, 2008.
	
	\bibitem{abraham78}
	R.~Abraham and J.~Marsden.
	\newblock {\em Foundations of Mechanics}.
	\newblock Benjamin/Cummings Publishing Company Reading, Massachusetts, 1978.
	
	\bibitem{CARIENA2007}J. Carie\~na, J. Clemente-Gallardo, \& G.  Marmo\newblock Reduction procedures in classical and quantum mechanics. \newblock {\em International Journal Of Geometric Methods In Modern Physics}. 4:1363-1403,  2007
	
	\bibitem{Neves_2006}A. Neves, L. Padilha, A. Fontes, E. Rodriguez, C. Cruz, L.  Barbosa,\& C. Cesar, 
	\newblock Analytical results for a Bessel function times Legendre polynomials class integrals. 
	\newblock {\em Journal Of Physics A: Mathematical And General}. 39(4):293-296, 2006
	
	\bibitem{evans}L. Evans, 
	\newblock \em{Partial Differential Equations}. 
	\newblock American Mathematical Society, 2010
	
\end{thebibliography}

\end{document}